%% file: main.tex
\documentclass[acmsmall,screen,nonacm]{acmart}

\AtBeginDocument{%
  }
\renewcommand\footnotetextcopyrightpermission[1]{}

\usepackage{enumitem}
\usepackage[table]{xcolor}
\usepackage{indentfirst}
\usepackage[most]{tcolorbox}
\usepackage{float}

\newcommand{\find}[1]{
\begin{tcolorbox}[leftrule=1mm,toprule=0mm,bottomrule=0mm,left=1pt,right=2pt,top=2pt,bottom=2pt]%
\em #1
\end{tcolorbox}
}

\begin{document}

\title{Beyond Coverage and Mutation: A Blind-Spot Analysis of LLM-Generated Unit Tests}

\author{Xiaoyu Shi}
\email{xiaoyu.shi@aalto.fi}
\orcid{0009-0001-0585-8167}
\affiliation{%
  \institution{Aalto University}
  \city{Espoo}
  \country{Finland}
}

\author{Wenqiang Luo}
\email{wenqialuo4-c@my.cityu.edu.hk}
\orcid{0009-0005-4171-2025}
\affiliation{%
  \institution{City University of Hong Kong}
  \city{Hong Kong}
  \country{China}
}

\author{Boyang Yang}
\email{yby@ysu.edu.cn}
\orcid{0000-0001-9270-730X}
\affiliation{%
  \institution{Yanshan University}
  \city{Hebei}
  \country{China}
}

\author{Haoye Tian}
\correspondingauthor
\email{haoye.tian@aalto.fi}
\orcid{0000-0002-8049-3997}
\affiliation{%
  \institution{Aalto University}
  \city{Espoo}
  \country{Finland}
}

\renewcommand{\shortauthors}{Shi et al.}

\begin{abstract}
Existing evaluation protocols for LLM-generated unit tests rely on aggregate metrics such as branch coverage and mutation score. These metrics quantify the extent to which a test suite exercises program behavior, but do not identify the behavioral obligations it systematically leaves unchecked.
We propose \textsc{BlindSpotEval}, an obligation-based evaluation framework aligned with the Arrange-Act-Assert pattern. The framework comprises three metrics, Boundary-Value Adequacy, Control-Flow Adequacy, and Oracle Strength, complemented by an assessment of execution reliability. We also introduce \textsc{ObligBench}, a complexity-stratified Java benchmark comprising 895 complex and 700 low-complexity functions with source-side obligation annotations.
Applied to 14,320 tests generated by four LLMs under four prompting strategies, \textsc{BlindSpotEval} reveals systematic blind spots that aggregate metrics do not distinguish. Before adequacy scoring, 19.7\% of tests fail to execute successfully, with failures ranging from compilation errors to runtime exceptions.
Among test failures, 15.6\% involve incorrect assertions, termed wrong-oracle failures. Such tests can contribute to code coverage despite failing, illustrating that coverage alone does not establish oracle correctness.
Among executable tests, LLMs readily cover directly inferable boundary cases but struggle with non-inferable ones. Coverage of \texttt{null}, \texttt{negative}, and \texttt{empty} arguments reaches only 34.8\%, 23.8\%, and 38.8\%, compared with average coverage above 93\% for booleans.
Exceptional control flow exhibits a similar pattern.Exceptional and fallback behavior is weakly checked, with CFA scores of 0.100 for catch blocks and 0.312 for default groups.
Across both dimensions, LLM-generated tests exhibit the {\em happy-path} bias observed in human-written tests, concentrating on normal inputs and expected flows while leaving unusual values and exceptional paths unchecked.
Oracle strength is high on average but uneven, with weaker assertions for object return property values and exceptions.
These blind spots also occur in low-complexity functions, suggesting that code complexity alone does not explain them.
\textsc{BlindSpotEval} provides construct-level adequacy profiles that identify under-tested obligation types and can guide targeted improvements beyond those supported by aggregate scores alone.
\end{abstract}

\maketitle

\input{sections/1.introduction}

\input{sections/2.related_work}

\input{sections/3.java_v2_benchmark_construction}

\input{sections/4.metrics}

\input{sections/5.experiment_settings}

\input{sections/6.results}

\input{sections/7.threats_and_conclusion}

\section*{Data Availability}
The ObligBench Benchmark, mutation tools, evaluation scripts, and results for BlindSpotEval are available at \url{https://github.com/rabbit0v0/Beyond-Coverage-and-Mutation-A-Blind-Spot-Analysis-of-LLM-Generated-Unit-Tests/}.

\bibliographystyle{ACM-Reference-Format}
\bibliography{references}

\end{document}

%% file: sections/1.introduction.tex
\section{Introduction}
\label{sec:introduction}

Software testing is fundamental to software quality, it provides a primary mechanism for developers to assess whether code behaves as intended and whether changes introduce regressions~\cite{ammann2016introduction,myers2011art}. As large language models (LLMs) increasingly support code generation and modification~\cite{chen2021codex}, automated unit test generation has become correspondingly important. Research on LLM-based test generation has advanced rapidly, producing a range of approaches. Zero-shot prompting generates tests directly from the function signature and body~\cite{chatunitest,testpilot,siddiq2024junit,tang2024chatgpt,yang2024evaluation}, feedback-based approaches iteratively refine tests using coverage or mutation signals~\cite{coveragent,dakhel2024mutap,yuan2024chattester}, and planning-based approaches first reason about the function's intent~\cite{gu2025panta}. Recent tools regularly produce tests that compile, run, and achieve high branch coverage and competitive mutation scores~\cite{testgeneval,chatunitest}.

High aggregate scores, however, do not necessarily indicate thorough testing. Prior work commonly evaluates LLM-generated tests using branch coverage, which measures the proportion of program branches executed~\cite{zhu1997software,fraser2011evosuite}, and mutation score, which measures the proportion of artificially injected faults detected~\cite{jia2011analysis,coles2016pit}. A replicability study found that these metrics correlate with fault detection only under restricted conditions~\cite{zhao2026coverage}. Although useful for summarizing test effectiveness, these metrics do not directly identify which obligations remain unchecked. %
For example, tests using only ordinary inputs may achieve the same line coverage as tests that also check boundary values whenever those inputs traverse the same statements. Similarly, a high mutation score indicates sensitivity to injected changes without necessarily establishing whether tests adequately check particular input boundaries, exceptional paths, or behavioral properties. These distinctions have practical consequences. Omitting \texttt{null} inputs may leave null-handling faults undetected, failing to exercise exception handlers may leave exception-swallowing behavior unchecked, and weak assertions may allow incorrect results to pass unnoticed.
Scores of 90\% mutation adequacy and 80\% branch coverage alone do not establish whether null inputs, exception paths, or object-state invariants have been adequately tested.

Unit tests are commonly structured according to the Arrange-Act-Assert (AAA) pattern~\cite{osherove2014art,meszaros2007xunit}. The \emph{Arrange} phase prepares the inputs, the \emph{Act} phase invokes the function, and the \emph{Assert} phase verifies the resulting behavior. This structure suggests three complementary evaluation dimensions, (1) tests should cover relevant input boundary values during Arrange; (2) tests should exercise relevant control-flow alternatives during Act; and (3) tests should verify the resulting behavior through their assertions. We refer to the behaviors that tests should exercise or verify, as identified from a function's implementation, as \emph{obligations}. Our aim is to identify systematic blind spots in how LLM-generated tests satisfy these obligations.

Existing approaches primarily assess generated tests through aggregate metrics. CoverAgent~\cite{coveragent} seeks to maximize branch coverage but provides no construct-level diagnosis of what remains unchecked; MuTAP~\cite{dakhel2024mutap} and ChatTester~\cite{yuan2024chattester} use mutation feedback to improve tests, with aggregate mutation scores serving as an evaluation measure. To the best of our knowledge, no existing framework jointly assesses LLM-generated tests at the obligation level across all three AAA dimensions.
We propose \textsc{BlindSpotEval}, an obligation-based evaluation framework aligned with the AAA pattern~\cite{osherove2014art,meszaros2007xunit}. For each function under test, the framework assesses which input boundaries, control-flow alternatives, and behavioral properties the generated tests exercise or verify.

Before assessing adequacy, we examine execution reliability through the following research question:

\textbf{RQ1 (Executability).}
\emph{To what extent do LLMs generate compilable and runnable tests, and what are the common failure modes?}
We measure execution success across multiple LLMs and prompting strategies and characterize failure modes to distinguish execution failures from adequacy gaps.

For tests that compile and execute successfully, we then investigate three dimensions of adequacy.

\textbf{RQ2 (Arrange).}
\emph{To what extent do LLM-generated tests cover the input boundary values relevant to a function?}
We propose \emph{Boundary-Value Adequacy} (BVA), which statically extracts per-parameter obligations and assesses whether the generated tests cover them.

\textbf{RQ3 (Act).}
\emph{To what extent do LLM-generated tests adequately exercise the control-flow alternatives within a function?}
We propose \emph{Control-Flow Adequacy} (CFA), which uses extreme-condition mutation to assess obligations associated with conditions, exception handlers, and switch cases.

\textbf{RQ4 (Assert).}
\emph{To what extent do LLM-generated tests verify a function's outputs and exception behavior?}
We propose \emph{Oracle Strength} (OS), which mutates return values and exception behavior and assesses whether the tests' assertions detect these changes. We exclude incidental kills caused by unexpected exceptions or crashes from the killed-mutant count, so that these failures do not inflate the estimated strength of the test oracles.

Finally, we examine whether the observed blind spots also occur in less complex code:

\textbf{RQ5 (Complexity).}
\emph{To what extent do the observed blind spots persist across levels of code complexity?}
We compare complex functions with a low-complexity control set to assess whether the identified gaps are concentrated in complex code or also occur in simpler functions.

To answer these questions we need a benchmark of real-world functions with known obligation profiles and a complexity-controlled comparison set. Existing benchmarks such as TestGenEval~\cite{testgeneval} evaluate test generation quality using aggregate metrics but do not provide obligation-level annotations together with a complexity-stratified comparison set. We therefore introduce ObligBench, a Java benchmark derived from The Stack v2~\cite{stackv2}. It comprises 895 complex functions under test, selected from the top 5\% by cyclomatic~\cite{mccabe1976} and cognitive~\cite{campbell2018cognitive} complexity, and a low-complexity control set of 700 functions. In the main experiment, we evaluate four LLMs (GPT-5.4, Gemma-4-31B, Gemini-3.6-flash, and Qwen3-Coder-30B) under four prompting strategies (zero-shot, intention-planning, mutation-feedback, and Panta~\cite{gu2025panta}) on the 895 complex functions, yielding 14,320 generated tests for RQ1-RQ4. For RQ5, we additionally evaluate tests generated for the 700-function low-complexity control set.

Our results reveal systematic adequacy blind spots that aggregate coverage and mutation scores do not distinguish. LLM-generated tests struggle most with inputs requiring semantic reasoning about the function's contract, as null inputs are covered in only 34.8\% of applicable cases, negative inputs in 23.8\%, and empty inputs in 38.8\%. The same happy-path bias appears in control-flow adequacy, with scores of 0.100 for catch blocks and 0.312 for default groups, as reaching these constructs requires deliberately out-of-domain inputs that LLMs tend to avoid. Taken together, these patterns show that LLMs exhibit the happy-path bias long observed in human-written test suites~\cite{barr2015oracle,beller2015developers} that, attention is concentrated on normal inputs and expected execution flows, while unusual values and exceptional paths are systematically omitted, forming blind spots that aggregate metrics do not surface. On the assertion side, oracle strength is high overall but weak for object returns when concrete default mutation can be made, where models default to coarse checks rather than field-level value assertions. Section~\ref{sec:results} presents the full results for all five research questions.

This paper contributes:
\begin{enumerate}
  \item \textbf{Evaluation framework.} \textsc{BlindSpotEval} combines an executability assessment with three AAA-aligned adequacy metrics (BVA, CFA, and OS). It characterizes execution failures before adequacy scoring and produces construct-level profiles that identify under-tested obligation types, supporting targeted improvements to generated tests.
  \item \textbf{Benchmark.} \textsc{ObligBench} comprises 895 complex and 700 low-complexity real-world Java functions under test (FUTs), with source-side obligation annotations and immutable-FUT constraints.
  \item \textbf{Empirical study.} An evaluation of four LLMs under four prompting strategies, comprising 14,320 generated tests on complex functions, characterizes execution failure modes and identifies recurring gaps in boundary-value coverage, exceptional control flow, and oracle strength. An additional study on 700 low-complexity functions indicates that these patterns are more pronounced in complex functions but also occur in simpler code. All artifacts are publicly available.
\end{enumerate}

%% file: sections/2.related_work.tex
\section{Background and Related Work}
\label{sec:related-work}

\emph{LLM-based test generation.}
Automated unit test generation has a long history of search-based~\cite{harman2001sbse,fraser2011evosuite,panichella2018automated}, random~\cite{pacheco2007randoop}, and symbolic-execution~\cite{king1976symbolic,cadar2008klee} approaches. Anand et al.~\cite{anand2013orchestrated} survey these methods before the emergence of LLM-based techniques. More recently, LLMs have become a prominent approach to test generation, both independently and in combination with established techniques. AthenaTest~\cite{tufano2020athenatest} formulates test generation as a sequence-to-sequence learning task, using pretraining followed by supervised fine-tuning on focal-method--test pairs. ChatUniTest~\cite{chatunitest} combines adaptive focal context with a generation-validation-repair workflow. TestPilot~\cite{testpilot} refines failing tests using execution errors, while CoverAgent~\cite{coveragent} iteratively generates and validates tests to improve code coverage. CodaMosa~\cite{lemieux2023codamosa} invokes an LLM to overcome coverage plateaus in search-based generation. MuTAP~\cite{dakhel2024mutap} uses surviving mutants to augment prompts for further test generation. ChatTester~\cite{yuan2024chattester} combines intention-based generation with iterative repair guided by compilation errors and additional code context. Panta~\cite{gu2025panta} integrates static control-flow analysis and dynamic coverage feedback to identify under-tested execution paths and guide subsequent generation. In an industrial setting, Alshahwan et al.~\cite{alshahwan2024meta} introduce TestGen-LLM, which extends existing tests and filters generated candidates for build success, execution reliability, and coverage improvement. These approaches commonly assess effectiveness through compilation or execution success, structural coverage, or aggregate mutation scores. \textsc{BlindSpotEval} complements these measures with obligation-level profiles that distinguish adequacy across input categories, control-flow constructs, and assertion targets.

\emph{Test adequacy criteria.}
Structural criteria, including line coverage, branch coverage, and modified condition/decision coverage (MC/DC), assess the execution of code elements and satisfaction of decision-related coverage requirements~\cite{ammann2016introduction,zhu1997software}. Mutation testing introduces artificial faults and assesses whether tests detect the resulting changes~\cite{demillo1978mutation,offutt2001mutation,jia2011analysis,coles2016pit,tip2025llmorpheus}. Although both approaches provide detailed information, their aggregate scores do not directly distinguish adequacy across input categories, construct types, and assertion targets. Prior work examines limitations of structural coverage~\cite{goodenough1975toward,inozemtseva2014coverage}, its relationship with mutation-based test generation~\cite{fraser2015achieving}, and the detection of real faults by generated tests~\cite{shamshiri2015autogenerated}. Other studies investigate whether mutation analysis reflects real-fault detection~\cite{andrews2005mutation,papadakis2018mutation}. For LLM-generated tests, a replicability study finds that the usefulness of coverage and mutation scores depends on the evaluation setting~\cite{zhao2026coverage}.

Boundary-value analysis emphasizes inputs at and around the boundaries of equivalence partitions~\cite{myers2011art,ammann2016introduction}. Category-partition testing similarly organizes the input space into categories and choices~\cite{ostrand1988category}. Our \emph{Boundary-Value Adequacy} (BVA) metric operationalizes these ideas through statically extracted per-parameter obligations. A boundary category receives credit only when the corresponding value is supplied at the relevant parameter position, allowing gaps to be attributed to specific categories and parameters.

Extreme mutation identifies pseudo-tested methods whose implementation logic can be removed without causing test failures~\cite{niedermayr2016extreme}. Vera-P{\'e}rez et al.~\cite{veraPerez2018descartes} extend this work through a broader empirical study of pseudo-tested methods. Our \emph{Control-Flow Adequacy} (CFA) metric instead targets individual control-flow constructs, attributing gaps to specific exception handlers, default branches, and throw paths.

For oracle evaluation, Barr et al.~\cite{barr2015oracle} survey the broader test oracle problem. Schuler and Zeller~\cite{schuler2013checked} introduce checked coverage, which uses dynamic slicing to identify executed statements that influence an oracle. Just et al.~\cite{just2014mutants} find an empirical relationship between mutant detection and real-fault detection, supporting mutation analysis as a proxy under the studied conditions. Our \emph{Oracle Strength} (OS) metric assesses assertion-based detection of mutations to return values and exception behavior. It gives incidental kills caused by unexpected exceptions or crashes no credits and reports scores by assertion subcategory, distinguishing oracle strength for object returns, exception behavior, and scalar values.

\noindent\emph{Benchmarks for automated test generation.}
Established benchmarks such as SF110~\cite{fraser2014large} and Defects4J~\cite{just2014defects4j} support the evaluation of test generation and fault detection and predate the recent use of LLMs for these tasks. More recent benchmarks draw on repositories and curated code corpora.
ULT~\cite{ult} provides curated function-level tasks involving complex Python functions, with an emphasis on mitigating test contamination. Its paired benchmark, PLT, supports comparisons with a setting in which reference tests are exposed. TestGenEval~\cite{testgeneval} provides code--test file pairs for full-file test generation and test completion, evaluated through execution success, coverage, and mutation scores. Other benchmarks address related objectives. EvalPlus~\cite{evalplus2023} augments HumanEval with more extensive test inputs to evaluate the functional correctness of generated code, while SWE-bench~\cite{swebench} targets repository-level issue resolution. These resources do not jointly provide source-side obligation annotations and a complexity-stratified comparison set for evaluating generated unit tests. \textsc{ObligBench} provides both through real-world Java functions drawn from The Stack v2~\cite{stackv2}, complexity-based selection, and source-side obligation annotations. Its evaluation protocol requires function bodies to remain unchanged so that adequacy is assessed against a fixed implementation.

%% file: sections/3.java_v2_benchmark_construction.tex
\section{ObligBench Construction}
\label{sec:java-v2-benchmark}

Existing test-generation benchmarks evaluate quality through aggregate metrics and do not jointly provide obligation-level annotations and a comparison set spanning different levels of complexity. To surface blind spots at the obligation level across all three AAA dimensions, we construct ObligBench from The Stack v2~\cite{stackv2}, a large licensed source-code corpus. We focus on Java because its explicit method signatures and mature analysis and testing ecosystem, including JaCoCo and PIT~\cite{coles2016pit}, support self-containment validation and obligation-level analysis.
Adapting the multi-stage construction approach of ULT~\cite{ult}, we select functions under test (FUTs) by complexity, and validate both high-complexity and low-complexity sets for self-containment and leakage before generation as discussed below.

{\em{Complexity-guided non-trivial FUT selection.}}
We select non-trivial FUTs using two complementary complexity metrics: (1) Cyclomatic complexity (CycC)~\cite{mccabe1976} counts independent execution paths, and (2) cognitive complexity (CogC)~\cite{campbell2018cognitive} additionally penalizes nesting and flow interruptions. Together, these metrics capture different aspects of control-flow complexity and provide a basis for selecting functions with varied test-generation demands.

We use the top 5\% under each metric to define two overlapping high-complexity views. In our hydrated Java sample, these thresholds are \(\mathrm{CycC} \geq 8\) and \(\mathrm{CogC} \geq 13\). Each view contains 700 tasks, of which 505 occur in both views. We retain the overlap and record view membership for each task so that subsequent analyses can account for the shared functions. The deduplicated union of 895 functions forms the \emph{treatment set} used in the main evaluation across LLMs and prompting strategies.
For RQ5, we additionally construct a 700-function \emph{control set} with scores below both thresholds (\(\mathrm{CycC} < 8\) and \(\mathrm{CogC} < 13\)). Both sets undergo self-containment validation and together constitute \textsc{ObligBench}. The control set is used only for the complexity comparison in RQ5.

{\em{Self-containment.}}
To enable test generation and execution without full repository context, we package each accepted function as an isolated executable Java task. The FUT body remains unchanged throughout this process, enforced by AST-token signature comparison against the original source. Each task passes a three-stage filter: parsing and type extraction, a smoke test that compiles the task and executes three distinct invocations, and an LLM wrapper-generation fallback for fixtures that fail to compile or execute.

{\em{Test leakage check.}}
Pre-existing tests in the corpus would give LLMs an unfair retrieval advantage~\cite{shi2023detecting}.  Following ULT~\cite{ult}, we check for existing tests associated with each FUT in the source corpus. We identify test-like files from the same Stack v2 repository and revision using path and content patterns, including \texttt{/test/}, \texttt{*Test.java}, \texttt{@Test}, and common JUnit or TestNG imports. Our check is a deterministic filter over the hydrated cache, and we flag a candidate only if it references the focal method, class, or package with sufficient heuristic evidence. In ObligBench, no likely corresponding Stack v2 tests were found for any of the 895 unique tasks.

Tables~\ref{tab:java-v2-complexity-profile} and~\ref{tab:java-v2-obligation-profile} summarize the complexity and obligation profiles of both control and treatment sets. Table~\ref{tab:java-v2-complexity-profile} shows that control FUTs have a median CycC of 1 and a median CogC of 1, while treatment FUTs reach a median CycC of 10 and CogC of 18. Table~\ref{tab:java-v2-obligation-profile} shows that treatment FUTs span a median of 9 branch-condition items across 35 lines, compared to 0 items and 3 lines in the control set. These profiles indicate that the treatment views contain more extensive control flow and more source-level branch obligations, providing a basis for examining adequacy gaps in complex functions.

\begin{table*}[t]
\centering
\scriptsize
\caption{Complexity profile of ObligBench treatment views and low-complexity control set.}
\label{tab:java-v2-complexity-profile}
\begin{tabular}{lrrrrrrrrr}
\hline
Split & $N$ & \multicolumn{4}{c}{CycC} & \multicolumn{4}{c}{CogC} \\
\cline{3-6}\cline{7-10}
 & & Min & Median & Mean & Max & Min & Median & Mean & Max \\
\hline
top5-CycC & 700 & 8 & 10.0 & 12.64 & 75 & 1 & 17.5 & 23.42 & 233 \\
top5-CogC & 700 & 1 & 9.0 & 11.68 & 75 & 13 & 18.0 & 25.05 & 233 \\
\hline
control    & 700 & 1 &  1.0 &  1.41 &  7 & 0 &  1.0 &  1.51 &  12 \\
\hline
\end{tabular}
\end{table*}

\begin{table*}[t]
\centering
\scriptsize
\caption{Source-obligation profile of ObligBench, percentages are relative to the split size. ``FUT lines'' and ``Branch items'' report medians per task.}
\label{tab:java-v2-obligation-profile}
\begin{tabular}{lrrrcc}
\hline
Split & $N$ & Exception & Branch tasks & FUT lines (med) & Branch items (med) \\
\hline
top5-CycC & 700 & 123 (17.6\%) & 689 (98.4\%) & 35 & 9 \\
top5-CogC & 700 & 150 (21.4\%) & 698 (99.7\%) & 35 & 9 \\
\hline
control    & 700 &  50 (7.1\%)  & 165 (23.6\%) &  3 & 0 \\
\hline
\end{tabular}
\end{table*}

%% file: sections/4.metrics.tex
\section{Evaluation Metrics}
\label{sec:evaluation-metrics}

Our research questions demand metrics that go beyond aggregate execution, coverage, and mutation scores. A test may execute successfully, cover branches, and kill mutants while still omitting important inputs, failing to check observable obligation, or asserting only coarse properties~\cite{barr2015oracle,hossain2023gaps,jahangirova2016oracle}. Aggregate metrics summarize test performance but do not directly identify which behavioral obligations remain unchecked. We therefore define a set of deterministic adequacy metrics that characterize complementary dimensions of generated-test behavior and produce construct-level profiles.

The proposed metrics follow the Arrange-Act-Assert (AAA) structure commonly used to organize unit tests~\cite{osherove2014art,meszaros2007xunit}. We measure \emph{Boundary-Value Adequacy} (Section~\ref{sec:boundary-value-adequacy}), which characterizes the inputs a test arranges; \emph{Control-Flow Adequacy} (Section~\ref{sec:control-flow-adequacy}), which characterizes the behavioral branches the test reaches and distinguishes; and \emph{Oracle Strength} (Section~\ref{sec:oracle-strength}), which whether tests check return values and exception behavior through targeted mutations. Each metric is
computed per generated test per FUT, and aggregate results are reported by model and prompting strategy. We refer to each test--FUT evaluation instance as a \emph{row}.

The three metrics use different measurement procedures.
BVA statically identifies selected input categories from the test source, independently of assertion quality.
CFA and OS use targeted mutations and test execution to assess detection, and they are not independent: an obligation the test never reaches
scores zero in CFA and is excluded from the OS denominator, so test suites with poor CFA may still have good OS. We report the eligible-obligation count
alongside OS summary scores. We first describe the shared obligation scoring pipeline, then define each metric in turn.

\subsection{Mutation-Based Obligation Scoring Pipeline}
\label{sec:measurement-infrastructure}

CFA and OS both score obligations by mutation: for each obligation, we construct mutants whose \emph{survival} implies the obligation was not exercised or not checked. This differs from classical mutation testing, which reports a single aggregate kill count treating equivalent mutants as noise~\cite{papadakis2019}. Here, we score by the stage where obligation belongs (arrange, act or assert), and the corresponding obligation must be killed, incidental kills do no contribute the scores. Two design choices distinguish our scoring: mutants are attributed to individual obligations and scored per-obligation, and unrepresentable obligations are excluded from adequacy denominators, with CFA additionally reporting a \emph{representativeness rate} to indicate tooling gaps.

The CFA and OS scoring pipeline separates into two stages: (1) \emph{collectors} generate mutants, compile and execute them, and write versioned artifacts, and (2) an \emph{analyzer} consumes those artifacts and computes scores without rerunning mutations.
Table~\ref{tab:eligibility-funnel} traces the Oracle Strength eligibility funnel from the 14{,}320 generated tests down to scored obligations, showing how population shrinks at each stage: test execution, obligation reachability, mutant viability, and oracle detection.

\begin{table}[t]
\centering
\caption{Eligibility funnel for Oracle Strength
(4 models $\times$ 4 prompts $\times$ 895 FUTs = 14{,}320 rows).
Obligations include return-value and exception behavior.}
\label{tab:eligibility-funnel}
\scriptsize
\begin{tabular}{@{}lr@{}}
\toprule
Stage & Rows / items \\
\midrule
Generated-test rows in set                      & 14{,}320 \\
\quad eligible after execution filtering        & 10{,}329 \\
\quad\quad non-private targets with mutation evidence
                                                & 6{,}682 \\
\midrule
Obligations found                               & 17{,}944 \\
\quad reached by the original test              & 14{,}338 \\
\quad\quad with an eligible, executed mutant    & 13{,}256 \\
\quad\quad\quad checked                         & 10{,}859 \\
\bottomrule
\end{tabular}
\end{table}

Not every mutant kill indicates that the test deliberately verified the changed behavior: a mutant may also be killed by an unexpected exception, which is an incidental kill. To exclude incidental kills, we classify assertion failures (\texttt{AssertionError} and its subtypes, including JUnit, OpenTest4J, and Mockito verification failures) as oracle kills, and all other escaping throwables as incidental. Incidental kills contribute zero to all scores. Moreover, an obligation the original test never reaches is excluded from OS and charged to CFA, and a mutant that could not be generated, failed to compile, or timed out reflects a tooling limitation and is excluded from adequacy denominators.

\subsection{Execution Reliability: Pre-Scoring Failure Classification}
\label{sec:execution-reliability}

We evaluate generated tests in a regression-testing setting, treating the original FUT as the reference implementation. Before adequacy scoring, we identify compilation failures, execution failures, and fixed placeholder outputs. These rows receive explicit skip statuses and are retained for reporting exclusion rates.

\emph{Compile failures} arise when the test cannot be compiled due to private-field access, unresolved symbols, type mismatches, checked-exception violations, or syntax errors.

\emph{Execution failures} arise when the test compiles but does not reach a clean end state. Subcategories include uncaught runtime exceptions and timeouts. We separately identify \emph{wrong-oracle failures}, in which an assertion specifies an expected result inconsistent with the reference implementation. Line and branch coverage may still record the code executed before the assertion fails, so coverage scores alone do not establish whether the generated test passes.

\emph{Generation failures} include Panta's fixed placeholder output, identified by its placeholder marker and small stub structure. This output contains a vacuous \texttt{assertTrue(true)} assertion and provides no meaningful oracle signal. Although it may compile and execute, it is excluded before adequacy scoring. Treating these as executable tests would silently inflate adequacy-rate denominators without contributing any oracle signal.

\subsection{Boundary-Value Adequacy: the Arrange Phase}
\label{sec:boundary-value-adequacy}

Boundary-value analysis emphasizes inputs at and around important partitions of the input domain. We define Boundary-Value Adequacy (BVA) as a parameter-specific input-space metric. A boundary category is credited only when it is used for the corresponding parameter in a generated call to the FUT.

\emph{Applicable boundary obligations.}
For each FUT, we infer applicable boundary obligations from its signature and simple source-level constraints.
There are 11 boundary categories: null, empty, false\_value, true\_value, zero, negative, below\_inferred\_bound, at\_inferred\_bound, above\_inferred\_bound, same\_inferred\_bound, and different\_inferred\_bound.

Signature-derived categories follow the declared type. Reference parameters receive a \texttt{null} obligation, strings and collections or arrays receive an \texttt{empty} obligation, numeric parameters receive \texttt{zero} and \texttt{negative} obligations, and boolean parameters receive obligations for both values.
The inferred-bound categories are source-derived, each direct parameter--literal comparison in the FUT body, such as \texttt{x < 5}, introduces \texttt{below}, \texttt{at}, and \texttt{above} obligations for that bound. Different inferred bounds are retained separately. String equality comparisons introduce an equality obligation for each inferred literal and a difference obligation requiring a value unequal to all inferred literals for that parameter. Boundary obligations are tracked separately for each parameter and each inferred bound. If a parameter is compared with multiple constants, each constant introduces its own boundary obligations.

\emph{Covered boundary obligations.}
We parse calls to the target method and classify each actual argument into boundary categories using deterministic rules. Coverage is parameter-specific: \texttt{f(-1, 0)} covers \texttt{negative} only for the first parameter and \texttt{zero} only for the second.

\emph{Argument resolution.}
Arguments are credited when they resolve to literals, simple local constants, or parameters of single-dispatch helper methods that pass their arguments directly to the FUT call. Values requiring deeper interprocedural reasoning are classified as unresolvable, section~\ref{sec:threats} discusses the implications of this restriction.

\emph{Score definition.}
Let \(B_{r,p}\) denote the applicable boundary obligations for parameter \(p\) in row \(r\), and let \(M_{r,p}\subseteq B_{r,p}\) denote those matched by generated calls. The parameter-level score is
\[
\mathrm{BVA}_{r,p}=
\frac{|M_{r,p}|}{|B_{r,p}|}.
\]
The row-level score averages over parameters with at least one applicable obligation,
\[
\mathrm{BVA}(r)=
\frac{1}{|P_r|}
\sum_{p\in P_r}\mathrm{BVA}_{r,p},
\qquad
P_r=\{p:B_{r,p}\neq\emptyset\}.
\]

\subsection{Control-Flow Adequacy: the Act Phase}
\label{sec:control-flow-adequacy}

We define Control-Flow Adequacy (CFA) to assesses whether tests adequately exercise control-flow alternatives, measuring whether generated tests check the behavioral alternatives selected by source-level control constructs. CFA measures whether changes in control-flow behavior is detectable by the generated tests, rather than execution coverage alone. A surviving mutant may indicate that the affected construct was not reached or that its changed behavior was not detected by the tests.

\emph{Applicable control-flow obligations.}
For each FUT, we extract \texttt{if} and \texttt{else-if} conditions, ternary conditions, switch \texttt{case} and \texttt{default} groups, \texttt{catch} blocks, and explicit \texttt{throw} statements. Consecutive switch labels sharing the same executable body are treated as one case group to avoid counting the same body multiple times.

\emph{Obligation-level control-flow mutants.}
Our approach is inspired by extreme mutation~\cite{niedermayr2016extreme} but targets individual obligations rather than entire methods.
Table~\ref{tab:custom-control-flow-mutants} lists the mutants and per-obligation scoring. An obligation the test never reaches receives 0. When the primary no-op or throw-removal mutant is non-viable, a type-aware fallback is attempted. Obligations without a viable mutation set are excluded.

\begin{table}[t]
\centering
\caption{Obligation-level control-flow mutants and per-obligation scoring.}
\label{tab:custom-control-flow-mutants}
\scriptsize
\begin{tabular}{lll}
\toprule
Construct & Obligation-level mutant & Obligation score \\
\midrule
\texttt{if}/\texttt{else-if} & condition \(\rightarrow\) \texttt{true};
                               \(\rightarrow\) \texttt{false} & 0, 0.5, 1 \\
ternary                      & condition \(\rightarrow\) \texttt{true};
                               \(\rightarrow\) \texttt{false} & 0, 0.5, 1 \\
\texttt{case} group          & executable case body \(\rightarrow\) no-op & 0 or 1 \\
\texttt{default} group       & executable default body \(\rightarrow\) no-op & 0 or 1 \\
\texttt{catch} block         & catch body \(\rightarrow\) no-op & 0 or 1 \\
exception path               & remove explicit \texttt{throw} & 0 or 1 \\
\bottomrule
\end{tabular}
\end{table}

\emph{Score definition.}
Let \(C_r\) be the applicable obligations for row \(r\) and
\(S_r\subseteq C_r\) the subset with at least one viable, executed
mutant. The representativeness rate and adequacy score are
\[
\mathrm{Repr}(r)=\frac{|S_r|}{|C_r|},
\qquad
\mathrm{CFA}(r)=\frac{1}{|S_r|}\sum_{i\in S_r}s_i.
\]

We report CFA alongside obligation counts and representativeness rates to distinguish high adequacy over a limited subset from high adequacy across most applicable obligations.

\subsection{Oracle Strength: the Assert Phase}
\label{sec:oracle-strength}

Branch coverage and standard mutation score both treat a test as adequate regardless of whether its assertions verify the exact resulting behavior. A test can achieve full branch coverage without asserting or asserting only that the function does not throw, leaving its actual result entirely unchecked. Even when meaningful assertions are present, kills can be incidental. A mutation may produce an unexpected exception and PIT records a kill, while the obligation was never intentionally checked. Oracle Strength (OS) addresses this blind spot by scoring each assertion obligation via targeted mutation that isolates changes to the asserted value.
For each supported obligation, we generate a mutant using a deterministic rule and observe the test outcome. The obligation is \emph{checked} when the mutant is killed through an oracle under the classification rules below. A surviving mutant or an incidental failure receives zero credit.

\emph{Oracle obligations.}
We extract obligations along two distinct constructs from each eligible row, which are the observable behaviors of FUTs. For each supported obligation, we generate a mutant using a deterministic rule and observe the test outcome. The obligation is \emph{checked} when the mutant is killed through an oracle under the classification rules below. A surviving mutant or an incidental failure receives zero credit.

\begin{itemize}[leftmargin=1.5em, itemsep=1pt, topsep=2pt, parsep=0pt]
\item {\em{Return-Obligation Mutation.}}
\label{sec:return-behavior}
We replace the returned expression using the first applicable rule in a fixed hierarchy, summarized in Table~\ref{tab:return-mutation-rules}.
The fixed selection procedure makes the probe reproducible.

For domain objects, the analysis checks whether the relevant state is observable through accessible fields, accessors, or an overridden \texttt{equals()} method when required by the replacement rule. Returns identified as having unobservable state are excluded from the oracle denominator.
Replacing an opaque object with \texttt{null} probes null sensitivity rather than object-content checking and is the weakest probe and tagged separately. We report results by obligation type and, for object returns, by mutation rule.

\item {\em{Exception-Obligation Mutation.}}
\label{sec:exception-behavior}
We replace the thrown exception type with a type-compatible alternative, respecting checked-exception constrains. Candidates handled identically by an enclosing \texttt{catch} clause are rejected.
For broad original types, such as \texttt{RuntimeException} or \texttt{Exception}, subtype replacements cannot be distinguished by assertions that accept the original type and its subtypes. We tag these obligations \texttt{subsumed\_by\_broad\_original} and exclude them from the headline score.
\end{itemize}

\begin{table}[t]
\centering
\caption{Return-Obligation mutation rules, applied in priority order.}
\label{tab:return-mutation-rules}
\scriptsize
\begin{tabular}{cll}
\toprule
Rule & Returned type or expression & Replacement \\
\midrule
1 & Boolean & negate literal; \texttt{!(expr)} \\
2 & Integral / \texttt{char} & $0\leftrightarrow1$; \texttt{expr\,+\,1} \\
3 & Floating-point & literals only \\
4 & String & \texttt{""} or sentinel; \texttt{expr\,+\,"MUTATED1"} \\
5 & Collection / map / array & empty $\leftrightarrow$ single-element instance \\
6 & Enum & next constant in declaration order \\
7 & Class token & bound-compatible alternative \\
8 & Literal \texttt{null} return & type-compatible non-null default \\
9 & Constructible domain object & defaults-instance (observability required) \\
10 & Opaque object & \texttt{null} (tagged \texttt{degenerate\_null}) \\
\bottomrule
\end{tabular}
\end{table}

Exception mutations require an additional outcome rule. When the injected exception type escapes the test, we classify the result as an oracle kill because the original exception was handled but its replacement was not. This provides evidence of type-sensitive exception handling, while a test that accepts a supertype covering both exceptions does not distinguish the replacement and records a survival, correctly reflecting the absence of type
discrimination.

Some tests catch and suppress failures raised by their own checks, allowing execution to complete despite a failed check. For example, calls to \texttt{fail()} or \texttt{throw new AssertionError()} may be enclosed in a handler for \texttt{AssertionError} or \texttt{Throwable} that does not rethrow the failure. We exclude these swallowed checks from the assertion analysis and record their count as a diagnostic. The rows remain in the pipeline, but the suppressed checks cannot contribute oracle credit.

\emph{Score definition.}
Let \(O_r^{\mathrm{sc}}\) denote the reached oracle obligations with an eligible, executed mutant after the exclusions above. For each obligation \(i\), define
\[
s_i=
\begin{cases}
1, & \text{if its mutant is killed by an oracle},\\
0, & \text{if its mutant survives or fails incidentally}.
\end{cases}
\]
The row-level score is
\[
\mathrm{OS}(r)=
\frac{1}{|O_r^{\mathrm{sc}}|}
\sum_{i\in O_r^{\mathrm{sc}}}s_i.
\]
Credit establishes detection of the selected mutation rather than exhaustive verification of the corresponding output.

\subsection{Metric Development and Validation}
\label{sec:metric-development}

The authors designed the metric definitions, scoring rules, and validation methodology. AI assistance was limited to script implementation. The authors inspected and refined the resulting scripts iteratively.

For BVA, we sampled up to 20 generated-test instances from each combination of boundary subtype, checked/unchecked status, and control/treatment set. This stratified procedure yielded 780 instances across eleven subtypes, with some strata containing fewer than 20 available examples. Two authors independently labeled these instances, achieving agreement with the classifier of \(\kappa=0.985\) and \(\kappa=0.972\), both within the almost-perfect agreement category of the Landis--Koch scale~\cite{landiskoch1977}.

For OS, two authors independently audited 100 assertion-mutation cases for mutant validity and correctness of the analyzer's outcome classification. Retaining \texttt{unsure} as a separate category, Gwet's AC1 was 0.94 for mutant validity and 0.87 for outcome classification~\cite{gwet2008agreement}.

%% file: sections/5.experiment_settings.tex
\section{Experimental Setup}
\label{sec:experimental-settings}

\subsection{Benchmark and Execution Protocol}
All experiments use \textsc{ObligBench} (Section~\ref{sec:java-v2-benchmark}), comprising 895 treatment functions under test (FUT) and
a 700-task low-complexity control set. Each generated test is evaluated in an isolated
Maven Java~17 task project. Tests are first compiled and executed, executable
tests are then evaluated using PIT mutation score and the three adequacy metrics defined in
Section~\ref{sec:evaluation-metrics}. Non-executable tests are retained for
failure analysis (Section~\ref{sec:rq1}) but excluded from adequacy-score
averages. The treatment set provides 14{,}320 test--FUT rows for RQ1-RQ4, while the control set is used only for the complexity comparison in RQ5.

\subsection{Prompting Strategies}
\label{sec:prompting-strategies}

We compare four generation strategies. All receive the same function-under-test
source code and benchmark metadata, and produce JUnit~5 test files under a
shared execution contract.

{\em{Zero-shot.}}
The zero-shot strategy generates a complete JUnit~5 test file from function
source and task metadata alone, without examples, planning, repair, or mutation
feedback. It serves as the simplest baseline, measuring what the model produces
from source code alone.

{\em{Intention-planning.}}
Inspired by plan-and-generate approaches~\cite{tufano2020athenatest,chatunitest,ryan2024codeaware},
intention-planning separates test \emph{planning} from \emph{implementation},
the model first produces a structured plan covering target behavior, inputs,
branch cases, and exception conditions, then generates the test file using
both the plan and source code. This makes intended obligations explicit before
code generation.

{\em{Mutation-feedback.}}
Mutation-feedback extends intention-planning with one feedback round, after the
initial test is executed, surviving PIT mutants are summarized as a compact
feedback object that the model uses to improve the
test~\cite{dakhel2024mutap,coveragent}. We bound the strategy to a single
feedback round to keep conditions comparable across models.\footnote{MuTAP~\cite{dakhel2024mutap} and ChatTester~\cite{yuan2024chattester} are natural candidates for direct comparison but could not be integrated, as their harnesses are tightly coupled to specific model APIs (GPT-3.5 and GPT-4) with fixed structured-output contracts incompatible with our pipeline. Reimplementing them would conflate harness design with prompt-strategy differences, we instead capture their mutation-evidence-as-feedback principle in our mutation-feedback strategy under a shared execution contract.}

{\em{Panta~\cite{gu2025panta}.}}
Panta is an external tool-assisted agentic baseline that manages iterative
generation and repair at the tool level, configured to one fix iteration per
task. It serves as an agentic repair-loop baseline and separates
the contribution of prompt strategy from that of tool-level integration.

\subsection{Model Suite}
\label{sec:model-suite}
We evaluate all four strategies using GPT-5.4 (OpenAI, proprietary large), Gemma-4-31B (RedHatAI,
open-weight mid), Gemini-3.6-flash (Google DeepMind, proprietary lightweight), and Qwen3-Coder-30B (Alibaba, open-weight code-specialist). The suite includes proprietary models from OpenAI and Google, an open-weight Gemma checkpoint distributed by RedHatAI, and an open-weight coding model from Alibaba's Qwen family. These models provide variation in model family, accessibility, and specialization.
For RQ5, we select GPT-5.4 and Gemma-4-31B as the highest and lowest performers in the suite, bracketing the capability range across the proprietary/open-weight divide. Qwen is excluded due to its elevated failure rate under guided strategies, while Gemini-3.6-flash, as a lightweight
intermediate-tier model, would not meaningfully extend this contrast.

\subsection{Decoding and Reproducibility}
All generation uses temperature~0. The 4 models $\times$ 4 strategies over 895 treatment functions yield \textbf{14{,}320 generated test rows}, or 3{,}580 rows per model.
Each result row records the model, strategy, task, raw response, extracted test code, execution status, PIT artifacts, and adequacy-metric artifacts. The analysis stage is deterministic and fully reproducible from those artifacts without rerunning mutations.

%% file: sections/6.results.tex
\section{Experimental Results}
\label{sec:results}

\subsection{RQ1: To What Extent Do LLMs Generate Compilable and Runnable Tests?}
\label{sec:rq1}

\noindent\textbf{[Objective and design]:}
Successful compilation and execution are prerequisites for the adequacy
analyses in RQ2--RQ4. This RQ therefore examine how often LLM-generated tests compile and run successfully, and characterizes the failures that occur.
The experiment covers 895 treatment FUTs, four LLMs, and four generation strategies, yielding 14{,}320 test--FUT rows (Section~\ref{sec:experimental-settings}).
We record two success rates, \emph{compilation success} and \emph{execution success}, and classify failed tests using the
categories defined in Section~\ref{sec:execution-reliability}. Furthermore, we record the generation failure specially for Panta when it produces a placeholder stub instead of a test.
Only successfully executed, non-placeholder tests proceed to adequacy analysis in RQ2-RQ4.

\noindent\textbf{[Results]:}
Of the 14{,}320 generated-test rows, 10{,}329 (72.1\%) are eligible for adequacy analysis. The remaining 3{,}991 rows comprise 1{,}437 compilation failures, 1{,}384 execution failures, 1{,}073 placeholder outputs, and 97 unclassified outcomes. These correspond to 10.0\%, 9.7\%, 7.5\%, and 0.7\% of all generated-test rows, demonstrating that LLMs are generally capable of generating compilable and runnable unit tests. Table~\ref{tab:failure-categories} presents the failure breakdown.

Panta placeholders form the largest individual category, accounting for 26.9\% of excluded rows. They contain a vacuous assertion and provide no meaningful test of the FUT. We identify them using the fixed placeholder marker and file-size threshold and exclude them before adequacy scoring.

Wrong-oracle failures account for 623 rows, or 15.6\% of excluded rows. Under our regression-testing assumption, these assertions specify expectations inconsistent with the reference implementation. Coverage tools may record the code executed before an assertion fails, so line or branch coverage alone does not reveal whether the generated expectations are consistent with that implementation. This distinction motivates reporting execution outcomes alongside coverage and adequacy scores.

\begin{table*}[t]
\centering
\caption{LLM-generated failed test breakdown. \emph{Left}: failure categories, percentages relative to the excluded rows. \emph{Right}: failure rate by strategy and model, percentages relative to the corresponding group size.}
\label{tab:failure-categories}
\scriptsize
\setlength{\tabcolsep}{4pt}
\begin{minipage}[t]{0.46\linewidth}
\vspace{0pt}
\centering
\begin{tabular}{@{}llrr@{}}
\toprule
Family & Category & Rows & \% \\
\midrule
Gen.\ failure & Panta placeholder stub & 1,073 & 26.9 \\
Execution & Uncaught exception    &  707 & 17.7 \\
          & Wrong oracle          &  623 & 15.6 \\
          & Timeout / hang        &   42 &  1.1 \\
          & Other                 &   12 &  0.3 \\
Compile   & Private access        &  446 & 11.2 \\
          & Unresolved symbol     &  432 & 10.8 \\
          & Syntax / other        &  403 & 10.1 \\
          & Type mismatch         &  104 &  2.6 \\
          & Checked exc.          &   52 &  1.3 \\
Unclassified &                    &   97 &  2.4 \\
\bottomrule
\end{tabular}
\end{minipage}%
\hfill%
\begin{minipage}[t]{0.48\linewidth}
\vspace{0pt}
\centering
\begin{tabular}{@{}lrrrl@{}}
\toprule
Group & Total & Failed & Rate & Dominant \\
\midrule
Zero-shot          & 3,580 & 1,171 & 32.7\% & Uncaught exc. \\
Intention-planning & 3,580 &   769 & 21.5\% & Wrong oracle \\
Mutation-feedback  & 3,580 &   951 & 26.6\% & Wrong oracle \\
Panta              & 3,580 & 1,100 & 30.7\% & Placeholder stub \\
\midrule
GPT-5.4            & 3,580 &   480 & 13.4\% & Uncaught exc. \\
Gemma-4-31B        & 3,580 &   628 & 17.5\% & Placeholder stub \\
Gemini-3.6-flash   & 3,580 &   546 & 15.3\% & Placeholder stub \\
Qwen3-Coder-30B    & 3,580 & 2,337 & 65.3\% & Placeholder stub \\
\bottomrule
\end{tabular}
\end{minipage}
\end{table*}

Failure rates vary across strategies and models (Table~\ref{tab:failure-categories}, right). Intention-planning has the lowest strategy-level failure rate (21.5\%), followed by mutation-feedback (26.6\%), Panta (30.7\%), and zero-shot (32.7\%). The additional mutation-feedback stage therefore does not yield a lower failure rate than intention-planning in this experiment.

Panta's failures have a distinct composition. Of its 1{,}100 excluded outputs, 1{,}073 are placeholders. The remaining 27 are unclassified outcomes, accounting for 0.8\% of all Panta rows. Panta's relatively low compilation and execution failure rates should therefore be considered together with its generation failures, which produce placeholder stubs.

Across models, GPT-5.4 has the lowest failure rate (13.4\%), followed by Gemini-3.6-flash (15.3\%) and Gemma-4-31B (17.5\%). Qwen3-Coder-30B has the highest rate (65.3\%), accounting for 58.6\% of all excluded rows, whose failures span placeholders, access or reflection errors, incorrect expectations, unresolved symbols or API misuse, and uncaught exceptions.

\find{{\bf [RQ-1]} \textbf{[Findings]:} (1) LLMs achieve 90.0\% compilation and 80.3\% execution success, and 72.1\% are eligible for
adequacy analysis, confirming baseline test-generation capability.   (2) Panta placeholder stubs (26.9\%) is the leading failure category, while Qwen3-Coder-30B has the highest failure rate (65.3\%). (3) Wrong-oracle failures (15.6\%) are invisible to coverage tools, representing a silent false-confidence signal. \textbf{[Insights]:} LLM-generated tests are able to fulfill the least requirements for unit testing. }

\subsection{RQ2: Which Input Boundary Categories Are Systematically Missed?}
\label{sec:rq2}

\noindent\textbf{[Objective and design]:}
Coverage and mutation scores reveal how much of the FUT a test suite reaches, but not whether its inputs include the boundary values most likely to expose faults, such as \texttt{null}, \texttt{negative}, or \texttt{empty} arguments.
This RQ therefore measures how well LLM-generated tests cover the boundary values each FUT is designed to handle, corresponding to the \emph{Arrange} phase of the AAA pattern.
From RQ2 to RQ4, we always evaluate the tests eligible after RQ1, using the same four LLMs under the same four generation strategies.
For each FUT, we extract parameter-specific boundary obligations (Section~\ref{sec:boundary-value-adequacy}) and check which of them the tests cover.
We report boundary-value adequacy (BVA) by boundary category, model, and prompting strategy.

\begin{figure}[htbp]
  \centering
  \includegraphics[width=.95\columnwidth]{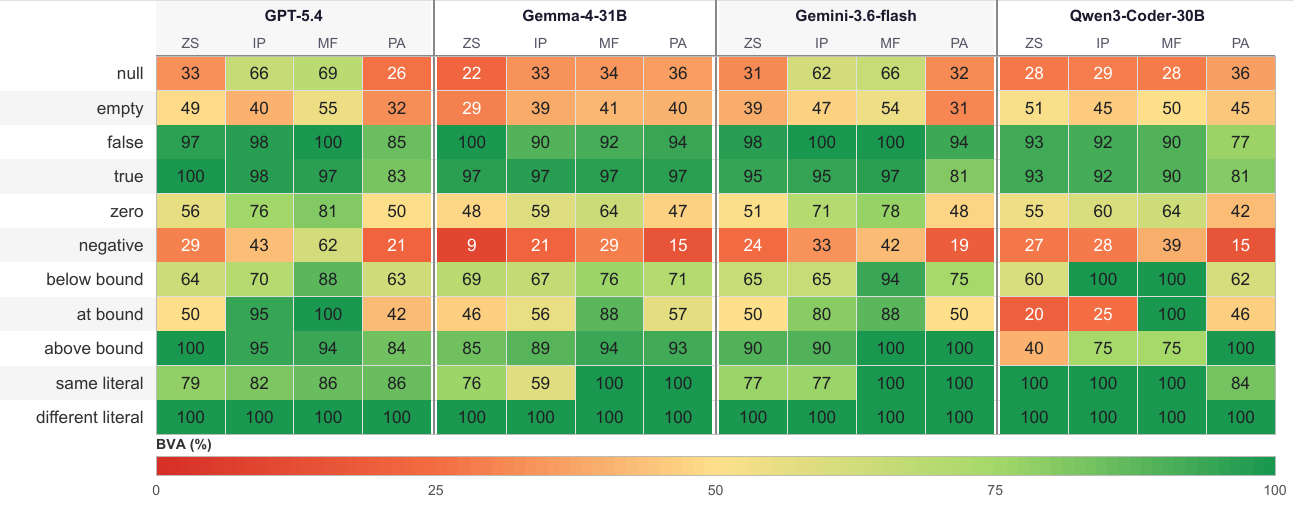}
  \caption{Boundary-Value Adequacy by category, model, and prompting strategy. ZS = zero-shot, IP = intention-planning, MF = mutation-feedback, PA = Panta.}
  \label{fig:bva-heatmap}
\end{figure}

\noindent\textbf{[Results]:}
Figure~\ref{fig:bva-heatmap} shows substantial variation across boundary categories.
Boolean obligations achieve high coverage, averaging 93.13\% for \texttt{true} and 93.75\% for \texttt{false}. The \texttt{different\_inferred\_bound} category reaches 100\% across all model--strategy combinations.
By contrast, \texttt{negative}, \texttt{null}, and \texttt{empty} obligations have coverage of only 23.8\%, 34.8\%, and 38.8\%. These remain the three lowest-coverage categories regardless of strategy or model, and together account for 17{,}200 of the 22{,}752 total applicable BVA obligations (75.6\%), their low coverage affects a substantial portion of the evaluated input space.

The BVA score asymmetry reveals a fundamental divide in how LLMs reason about test inputs, the generated tests more consistently include inputs with explicit alternatives or source-level cues.
For high-coverage categories, there is usually straightforward specification in the source code about what to test. \texttt{Boolean} boundary values are simply true and false, while the \texttt{different\_inferred\_bound} is derived from string literals appearing in equality comparisons.
However, the low-coverage categories require explicit selection of null references, negative numbers, or empty values, and non-trivial reasoning is required for LLMs to understand what boundary value needs to be covered. Although their applicability can be identified from parameter types, these input boundaries do not appear explicitly in the implementation. In our result, \texttt{null} handling is not specified by the declared type alone, \texttt{negative} requires hypothesising a domain violation on numeric inputs, and \texttt{empty} requires actively constructing a degenerate collection or string rather than the non-empty defaults models prefer. Their low coverage is consistent with a preference for ordinary inputs over boundary and exceptional cases.

Boundary-level analysis also distinguishes inputs that structural coverage may group together. Consider the guard \texttt{if (obj == null || obj.isEmpty())}. Tests using an empty and a non-empty string can exercise both outcomes of the \texttt{if} statement without supplying \texttt{null}. BVA records the missing null obligation separately from the covered empty-string obligation. This parameter-specific distinction is not expressed by an aggregate branch-coverage or mutation score.

\find{{\bf [RQ-2]} \textbf{[Findings]:} (1) LLM-generated tests get near-ceiling scores in boundary categories directly inferrable from source code, like boolean ($\sim$93\%) and \texttt{different\_literal} (100\%). (2) LLM-generated tests perform poorly across all models and strategies when the boundary value is not directly inferrable from source code, for example \texttt{null} (0.348), \texttt{negative} (0.238), and \texttt{empty} (0.388).  \textbf{[Insights]:} LLMs appear to struggle with reasoning about boundary cases whose relevance is not explicit in the method signature or source code. }

\subsection{RQ3: Which Control-Flow Constructs Are Systematically Blind Spots?}
\label{sec:rq3}

\noindent\textbf{[Objective and design]:}
Aggregate branch coverage reports how many branches a test suite reaches, but not which kinds of behavior remain unexercised, such as exception-handling paths or default cases.
This RQ therefore measures how thoroughly LLM-generated tests exercise the control-flow alternatives, corresponding to the \emph{Act} phase of the AAA pattern.
We evaluate eligible tests using the obligation-level mutations defined in Section~\ref{sec:control-flow-adequacy} and report CFA by construct, model, and prompting strategy. We also compare representativeness and detection scores with PIT-based measurements to assess the additional construct-level information provided by our operators.

\noindent\textbf{[Results]:}
Table~\ref{tab:cfa-heatmap} compares our proposed mutation operator and PIT on \emph{Representative rate} and \emph{Adequacy score}. The table reveals differences between the approaches.
First, the our mutation operators represent substantially more obligations for several constructs, particularly explicit \texttt{throw} statements. Second, PIT kill rates far exceed our CFA adequacy scores, making a PIT-only evaluation systematically overconfident. Since PIT represents a smaller subset of obligations, its scores may omit inadequately tested constructs that our operators include.

\begin{table}[htbp]
\centering
\caption{Control-flow adequacy by construct. Representativeness is the fraction of obligations with a generated mutant and adequacy score is the fraction killed, each compared across our obligation-level operator (CFA) and PIT. The PIT kill value for explicit \texttt{throw} is omitted because representativeness is below 2\%.}
\label{tab:cfa-heatmap}
\scriptsize
\setlength{\tabcolsep}{4pt}
\begin{tabular}{@{}lrrrrr}
\toprule
 & & \multicolumn{2}{c}{Representativeness rate} & \multicolumn{2}{c}{Adequacy score} \\
\cmidrule(lr){3-4}\cmidrule(lr){5-6}
Construct & Items & CFA Repr. & PIT Repr. & CFA Score & PIT Kill \\
\midrule
\texttt{catch}            &  3,257 & \cellcolor{green!25}0.849 & \cellcolor{yellow!35}0.453 & \cellcolor{red!25}0.100    & \cellcolor{yellow!35}0.445 \\
\texttt{default}          &  1,500 & \cellcolor{green!25}0.781 & \cellcolor{yellow!35}0.447 & \cellcolor{orange!25}0.312  & \cellcolor{yellow!35}0.449 \\
\texttt{if}               & 43,826 & \cellcolor{green!25}0.990 & \cellcolor{green!25}0.930  & \cellcolor{yellow!35}0.448  & \cellcolor{green!25}0.927  \\
\texttt{case}             & 13,797 & \cellcolor{green!25}0.987 & \cellcolor{yellow!35}0.593 & \cellcolor{yellow!35}0.440  & \cellcolor{yellow!35}0.591 \\
\texttt{else-if}          &  6,081 & \cellcolor{green!25}0.995 & \cellcolor{green!25}0.953  & \cellcolor{yellow!35}0.519  & \cellcolor{green!25}0.948  \\
\texttt{throw} (explicit) &  3,850 & \cellcolor{green!25}0.982 & \cellcolor{red!25}0.014    & \cellcolor{yellow!35}0.475  & ---                        \\
\texttt{ternary}          &  4,581 & \cellcolor{green!25}0.931 & \cellcolor{green!25}0.931  & \cellcolor{yellow!35}0.520  & \cellcolor{green!25}0.933  \\
\bottomrule
\end{tabular}
\end{table}

\begin{figure}[htbp]
  \centering
  \includegraphics[width=\columnwidth]{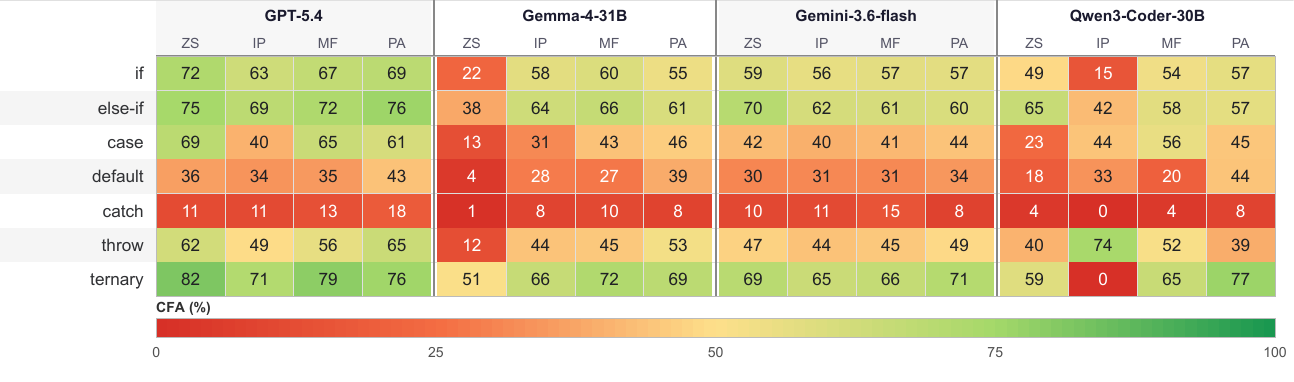}
  \caption{Control-flow Adequacy by construct and prompting strategy.
    ZS = zero-shot, IP = intention-planning, MF = mutation-feedback, PA = Panta.
    }
  \label{fig:cfa-heatmap}
\end{figure}

\emph{Representativeness across constructs.}
CFA Repr.\ and PIT Repr.\ expose how well each tool can generate mutants for obligations.
Our obligation-level operators achieve 78.1--99.5\% representativeness across constructs, while the configured PIT operators represent only 44.7--45.3\% of \texttt{catch} and \texttt{default} obligations and 1.4\% of explicit \texttt{throw} obligations. Our throw-removal operator raises representativeness to 98.2\%, allowing us to assess obligations largely unrepresented by PIT. Their CFA score of 0.475 reveals that many remain inadequately checked while  while PIT omits them.

\emph{Adequacy score under different mutation probes.}
PIT Kill scores exceed CFA scores across the reported constructs. For \texttt{if}, \texttt{else-if}, and ternary expressions, the difference may largely reflect the scoring rules. CFA assigns 0.5 when only one of the forced-true and forced-false mutants is killed, reserving full credit for both killed. A PIT-based indicator that credits an obligation whenever any matched mutant is killed does not distinguish partial from complete coverage of these alternatives.

For other constructs, differences in the represented obligation populations may also contribute. PIT represents fewer \texttt{catch}, \texttt{default}, and \texttt{case} obligations, so its adequacy scores may be calculated over smaller and potentially easier-to-detect subsets. Higher scores on these subsets therefore need not indicate stronger adequacy across the full obligation population. The comparison highlights the importance of interpreting adequacy scores alongside their denominators and representativeness rates.

\emph{Persistent gaps in exceptional and fallback behavior.}
The lowest CFA scores occur for \texttt{catch} blocks (0.100) and \texttt{default} groups (0.312), despite their relatively high representativeness under the targeted operators. Figure~\ref{fig:cfa-heatmap} shows that these weaknesses recur across models and prompting strategies. Adequately testing exception handlers may require constructing an exceptional execution state or inducing a dependency failure. Testing a switch default requires a selector value outside the explicit case labels. The low scores are consistent with a preference for ordinary inputs and expected execution flows, similar to the boundary-value omissions observed in RQ2.

\find{{\bf [RQ-3]} \textbf{[Findings]:} (1) Our proposed obligation-level control-flow mutation achieve 78.1--99.5\% representativeness, with a particularly large increase for explicit throw obligations compared with the configured PIT analysis. (2) PIT overstates adequacy on \texttt{ternary}/\texttt{if}/\texttt{else-if} (0.927--0.948) due to inability to score true and false outcomes separately. (3) Our operator reveals LLM-generated tests have poor coverage on \texttt{catch} (0.100) and \texttt{default} (0.312), while performing relatively good in other constructs. \textbf{[Insights]:} (1) Our obligation-level mutation reveals unexercised behavior that PIT cannot measure.  (2) LLMs bias toward happy-path inputs, systematically missing out-of-domain conditions such as \texttt{catch} and \texttt{default}.}

\subsection{RQ4: How Strongly Does the Test Oracle Check Return and Exception
Behaviour?}
\label{sec:rq4}

\noindent\textbf{[Objective and design]:}
Coverage score showing a return or exception line is reached does not establish that its behavior is meaningfully checked. Assertions may constrain only coarse properties, and unexpected execution failures can receive mutant-kill credit incidentally.
This RQ therefore measures how effectively LLM-generated tests check the FUT's return values and exception behaviors, corresponding to the \emph{Assert} phase of the AAA pattern.
For each FUT, we mutate its return values and exception obligations (Section~\ref{sec:oracle-strength}) and check whether the tests' assertions detect the change, excluding incidental kills from unexpected throws and crashes. We report OS over reached obligations with eligible mutants, broken down by return category and exception behavior across models and prompting strategies.

\begin{figure}[htbp]
  \centering
  \includegraphics[width=\columnwidth]{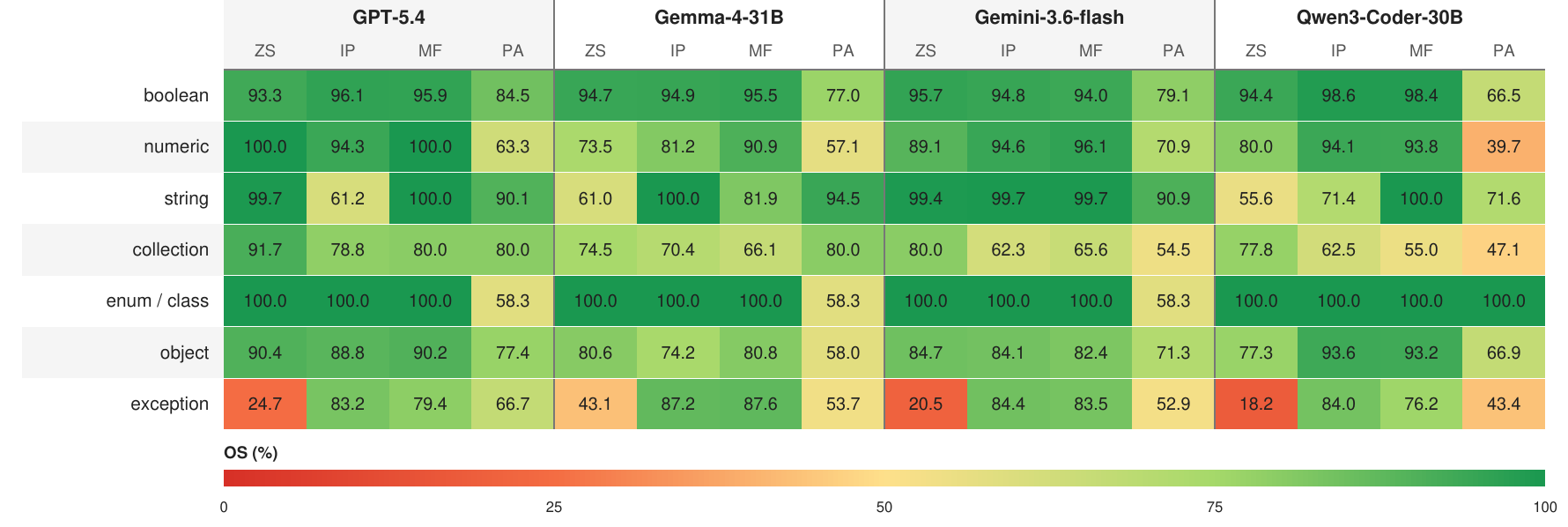}
  \caption{Oracle Strength by return category, model, and prompting strategy.
    ZS = zero-shot, IP = intention-planning, MF = mutation-feedback, PA = Panta.}
  \label{fig:os-return-type-heatmap}
\end{figure}

\noindent\textbf{[Results]:}
\noindent\emph{Return obligations.}
Figure~\ref{fig:os-return-type-heatmap} shows that boolean obligations score highest at 0.899, followed by enum/class tokens at 0.868 and strings at 0.862. Object-category and numeric obligations score 0.800 and 0.781, while collections score lowest at 0.710.
The differences may reflect the properties that assertions need to check. Boolean, enum/class-token, and string results can often be checked directly through some limited set of value or equality assertions. Collection results may require checks on size, membership, or contents, and numeric perturbations can survive coarse assertions, such as positivity or nonzero checks.

Object-category OS further varies with the property probed. Table~\ref{tab:os-object-rules} reports the four object-category mutation rules with scorable treatment obligations.
Null-compatible replacement scores 0.912, followed by opaque-object null substitution at 0.827 and default-constructor replacement at 0.766.
Concrete-constructor-defaults mutations score only 0.325. These mutations replace constructor arguments with default values, probing whether tests distinguish the resulting object state. Their low adequacy rate suggests that generated tests struggle to meaningfully check object contents beyond coarser properties such as null.

\begin{table}[t]
\centering
\caption{Oracle Strength for object-category mutation rules with scorable treatment obligations.}
\label{tab:os-object-rules}
\scriptsize
\begin{tabular}{@{}lrrr@{}}
\toprule
Mutation rule & Scorable & Checked & Score \\
\midrule
Null-compatible replacement      &   442 &   403 & 0.912 \\
Domain obj.: default constructor &   445 &   341 & 0.766 \\
Opaque: degenerate null          & 1{,}620 & 1{,}339 & 0.827 \\
Domain obj.: concrete defaults   &   163 &    53 & 0.325 \\
\bottomrule
\end{tabular}
\end{table}

\noindent\emph{Exception obligations.}
Exception obligations score 0.625, slightly below return obligations, average 0.842.
Exception-type substitution requires tests to distinguish the original exception from its replacement. Broad assertions, such as \texttt{assertThrows(Exception.class, ...)}, may accept both types and therefore leave the obligation unchecked under our criterion. The lower score suggests that generated tests less consistently check specific exception behavior.

This score applies only to reached obligations with eligible mutants. Exception obligations not reached by the original test are excluded from the OS denominator, since their oracle strength cannot be assessed without first exercising the corresponding behavior.

\find{{\bf [RQ-4]} \textbf{[Findings]:} (1) Return-category OS ranges from 0.710 for collections to 0.899 for booleans. Object-category OS is 0.800, but concrete-constructor-defaults obligations score only 0.325. (2) Reached and eligible exception obligations score 0.625, with substantial variation across generation strategies. \textbf{[Insights]:} (1) LLMs' capability to generate meaningful assertions for return value vary on type. Object returns require field-level assertions that LLMs rarely produce, defaulting to coarse checks. (2) The improvements in out-of-domain coverage may help improve effective exception OS. }

\subsection{RQ5: Does Complexity Impact the Systematic Weaknesses?}
\label{sec:rq5}

\noindent\textbf{[Objective and design]:}
The preceding results identify recurring adequacy gaps in complex functions. This RQ examines whether similar gaps also occur in simpler code.
We run test generation on a 700-function low-complexity control set with GPT-5.4 and Gemma-4-31B selected in Section~\ref{sec:model-suite} using the same four generation strategies and adequacy analyses. Control functions fall below both complexity thresholds defined in Section~\ref{sec:java-v2-benchmark}.
We compare BVA, CFA, and OS between the complex and control sets per obligation category, if a category remains weak in both sets, complexity alone is unlikely to explain the gap.

\begin{figure}[ht]
\centering
\includegraphics[width=.95\linewidth,trim=0 0 0 25,clip]{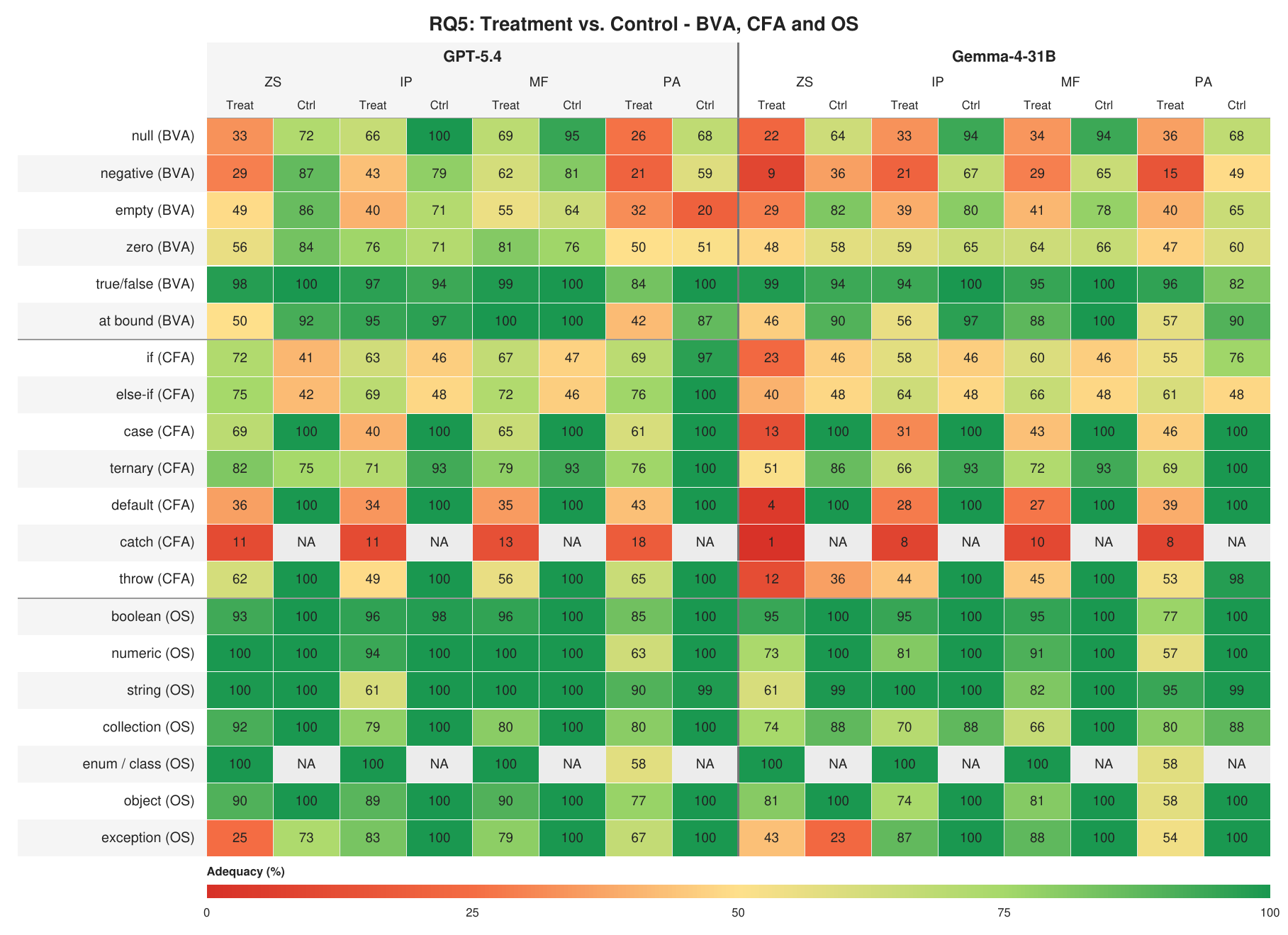}
\caption{Adequacy of Control and Treatment by subcategory, model, and prompting strategy. ZS = zero-shot, IP = intention-planning, MF = mutation-feedback, PA = Panta.}
\label{fig:rq5-heatmap}
\end{figure}

\noindent\textbf{[Results]:}
We present subcategory adequacy scores for both models across all four strategies in Figure~\ref{fig:rq5-heatmap}.

\emph{Boundary-value adequacy.}
Figure~\ref{fig:rq5-heatmap} shows that several patterns from RQ2 persist across complexity levels. Boolean values and \texttt{at\_inferred\_bound} obligations receive high coverage in both sets, while negative and empty inputs remain among the least covered categories across models and strategies. These recurring gaps show that the weaknesses are not confined to complex functions and suggest that complexity alone does not explain them.

Null coverage is higher in the control set, indicating that some boundary categories are better covered in simpler functions. However, the control set contains only 74 applicable nullable parameters, limiting how broadly this difference can be interpreted. Overall, the comparison supports the persistence of specific boundary-value blind spots across complexity levels, rather than a uniform effect of complexity on all categories.

\emph{Control-flow adequacy.}
The control set contains no applicable catch-block obligations, so the corresponding cells provide no basis for comparison. For \texttt{if} and \texttt{else-if}, several control-set configurations have low CFA scores despite the functions' low structural complexity. This indicates that simpler control flow does not necessarily receive adequate behavioral checking.

Some control-flow comparisons are based on small obligation populations. The reported \texttt{else-if} comparisons include only 26-29 scorable items, while several other construct categories contain fewer than ten. Scores in these categories can therefore change substantially with the outcomes of a few obligations and should be interpreted cautiously.

\emph{Oracle strength.}
Return categories with high treatment OS remain strong in control. However, control provides limited evidence for the remaining categories. Object OS covers only 87 scorable obligations from 11 methods under one null-replacement rule, while the lower-performing treatment rules are absent.
Enum/class has no scorable control obligations, and exception scores cover only 13 distinct methods.
Thus, the comparison confirms strong checking in the well-supported return categories but no clear conclusion can be drawn for the weak categories.

\find{{\bf [RQ-5]} \textbf{[Findings]:} (1) Negative and empty inputs remain weak BVA categories in
both treatment and control functions.  (2) Low-complexity FUTs have fewer obligations to check, limiting interpretability.  \textbf{[Insights]:} (1) For input boundary values present in both sets, the same blind-spot pattern holds across complexity levels, confirming that the blind spot patterns we found are a property of LLM test generation, not of code complexity alone.  }

%% file: sections/7.threats_and_conclusion.tex
\section{Threats to Validity}
\label{sec:threats}

{\em{Construct validity.}}
BVA uses deterministic static analysis, and 29.3\% of the target-call arguments cannot be resolved by the supported analysis rules. We report this proportion explicitly to distinguish the scope of input resolution from the adequacy of the generated tests. The manual validation described in Section~\ref{sec:metric-development} assesses the reliability of the resulting classifications.

{\em{Internal validity.}}
The choice of CFA mutation operators may influence the measured adequacy of different control-flow constructs. We address this through deterministic, construct-specific transformations applied
uniformly across models, strategies, and complexity sets. We also report representativeness rates alongside adequacy scores, distinguishing limitations in mutant availability from failures to detect represented behavioral changes.

{\em{External validity.}}
The blind-spot patterns are consistent across all four models from different families and four different strategies, reducing the risk that any single model or prompt design drives the results.
Results may differ for very small models, domain-specialized code models, or prompt designs not studied here. All generation uses temperature~0, results under non-zero temperature may show
higher variance.

\section{Conclusion}
\label{sec:conclusion}

We propose \textsc{BlindSpotEval}, an obligation-based evaluation framework for LLM-generated unit tests, evaluating tests through Boundary-Value Adequacy, Control-Flow Adequacy, and Oracle Strength, complemented by execution reliability and failure-mode analysis.
We also introduce \textsc{ObligBench}, a complexity-stratified Java benchmark of 895 complex and 700 low complexity FUTs with source-side obligation annotations.
Applied to 14{,}320 generated tests across four LLMs and four prompting strategies on ObligBench treatment set, \textsc{BlindSpotEval} reveals systematic blind spots that aggregate coverage and mutation scores do not distinguish.

Execution reliability is the first barrier, with 72.1\% of generated-test rows eligible for adequacy analysis. Wrong-oracle failures account for 15.6\% of excluded rows, demonstrating that executed code can receive coverage credit even when generated assertions fail.
Beyond execution, the dominant weakness is out-of-domain value reasoning and bias to happy path.
Boolean and inferred-bound BVA categories are near-ceiling across all models and
strategies, whereas \texttt{null}, \texttt{negative}, and \texttt{empty} obligations reach only 34.8\%, 23.8\%, and 38.8\%, for which the boundary values are not directly inferrable from the source code.
These omissions, together with low CFA for \texttt{catch} blocks (0.100) and \texttt{default} groups (0.312), are consistent with a happy-path preference in generated tests.

Obligation-based control-flow mutations broaden construct-level evaluation, particularly for explicit throw obligations, which achieve 98.2\% representativeness compared with 1.4\% under the configured PIT analysis. Separate forced-outcome mutations also distinguish partial checking of conditional alternatives that an any-kill indicator does not express.
Oracle Strength further distinguishes meaningful checking from incidental execution failures. Although return obligations score 0.842 overall, concrete-constructor-defaults obligations score only 0.325, exposing a pronounced gap in detecting changes to constructed object state. Exception obligations score 0.625, further distinguishing the ability to reach exceptional behavior from the ability to check it.

The low-complexity comparison shows that selected boundary-value and control-flow blind spots also occur in simpler functions, while well-supported OS categories remain strongly checked across both complexity sets. Across the study, obligation-level profiles identify the input categories, control-flow constructs, and observable behaviors that require stronger testing.

Together, these findings motivate shifting from aggregate coverage and mutation scores to construct-level adequacy profiles that reveal \emph{where} generated tests leave behavioral obligations unchecked and guide targeted improvements.

%% file: references.bib
@inproceedings{gu2025panta,
  author    = {Sijia Gu and Noor Nashid and Ali Mesbah},
  title     = {{LLM} Test Generation via Iterative Hybrid Program Analysis},
  booktitle = {Proceedings of the 48th International Conference on Software Engineering (ICSE)},
  pages     = {655--666},
  year      = {2026},
  doi       = {10.1145/3744916.3764553},
  eprint    = {2503.13580},
  archivePrefix = {arXiv},
}

@article{ult,
  author    = {Dong Huang and
               Jie M. Zhang and
               Mark Harman and
               Qianru Zhang and
               Mingzhe Du and
               See-Kiong Ng},
  title     = {Benchmarking {LLMs} for Unit Test Generation from Real-World Functions},
  journal   = {arXiv preprint arXiv:2508.00408},
  year      = {2025},
}

@inproceedings{testgeneval,
  author    = {Kush Jain and Gabriel Synnaeve and Baptiste Rozi{\`e}re},
  title     = {{TestGenEval}: A Real World Unit Test Generation and Test Completion Benchmark},
  booktitle = {Proceedings of the 13th International Conference on Learning Representations (ICLR)},
  year      = {2025},
}

@inproceedings{swebench,
  author    = {Carlos E. Jimenez and
               John Yang and
               Alexander Wettig and
               Shunyu Yao and
               Kexin Pei and
               Ofir Press and
               Karthik Narasimhan},
  title     = {{SWE-bench}: Can Language Models Resolve Real-World {GitHub} Issues?},
  booktitle = {Proceedings of the 12th International Conference on Learning
               Representations (ICLR)},
  year      = {2024},
}

@article{stackv2,
  author    = {Anton Lozhkov and Raymond Li and Loubna Ben Allal and Federico Cassano and Joel Lamy-Poirier and Nouamane Tazi and Ao Tang and Dmytro Pykhtar and Jiawei Liu and Yuxiang Wei and others},
  title     = {StarCoder 2 and The Stack v2: The Next Generation},
  journal   = {arXiv preprint arXiv:2402.19173},
  year      = {2024},
}

@article{fraser2014large,
  author    = {Gordon Fraser and
               Andrea Arcuri},
  title     = {A Large-Scale Evaluation of Automated Unit Test Generation Using
               {EvoSuite}},
  journal   = {ACM Transactions on Software Engineering and Methodology},
  volume    = {24},
  number    = {2},
  pages     = {8:1--8:42},
  year      = {2014},
  doi       = {10.1145/2685612},
}

@inproceedings{chatunitest,
  author    = {Yinghao Chen and
               Zehao Hu and
               Chen Zhi and
               Junxiao Han and
               Shuiguang Deng and
               Jianwei Yin},
  title     = {{ChatUniTest}: A Framework for {LLM}-Based Test Generation},
  booktitle = {Companion Proceedings of the 32nd ACM International Conference on
               the Foundations of Software Engineering (FSE 2024 Demonstrations)},
  pages     = {572--576},
  year      = {2024},
}

@article{testpilot,
  author    = {Max Sch{\"{a}}fer and
               Sarah Nadi and
               Aryaz Eghbali and
               Frank Tip},
  title     = {An Empirical Evaluation of Using Large Language Models for Automated
               Unit Test Generation},
  journal   = {IEEE Transactions on Software Engineering},
  volume    = {50},
  number    = {1},
  pages     = {85--105},
  year      = {2024},
  doi       = {10.1109/TSE.2023.3334955},
}

@misc{coveragent,
  author       = {{CodiumAI}},
  title        = {{CoverAgent}: An {AI}-Powered Tool for Automated Test Generation and Code Coverage Enhancement},
  year         = {2024},
  howpublished = {\url{https://github.com/qodo-ai/qodo-cover}},
}

@article{tufano2020athenatest,
  author    = {Michele Tufano and Dawn Drain and Alexey Svyatkovskiy and Shao Kun Deng and Neel Sundaresan},
  title     = {Unit Test Case Generation with Transformers},
  journal   = {arXiv preprint arXiv:2009.05617},
  year      = {2020},
}

@article{papadakis2019,
  author    = {Mike Papadakis and
               Marinos Kintis and
               Jie Zhang and
               Yue Jia and
               Yves Le Traon and
               Mark Harman},
  title     = {Mutation Testing Advances: An Analysis and Survey},
  journal   = {Advances in Computers},
  volume    = {112},
  pages     = {275--378},
  year      = {2019},
  doi       = {10.1016/bs.adcom.2018.03.015},
}

@article{zhao2026coverage,
  author    = {Junda Zhao and
               Shurui Zhou and
               Eldan Cohen},
  title     = {Do Coverage and Mutation Scores of {LLM}-Generated Test Suites
               Correlate with Their Effectiveness? (Replicability Study)},
  journal   = {Proceedings of the ACM on Software Engineering},
  volume    = {3},
  number    = {ISSTA},
  pages     = {30--53},
  year      = {2026},
  doi       = {10.1145/3832093},
}

@inproceedings{fraser2011evosuite,
  author    = {Gordon Fraser and
               Andrea Arcuri},
  title     = {{EvoSuite}: Automatic Test Suite Generation for Object-Oriented
               Software},
  booktitle = {Proceedings of the 19th ACM SIGSOFT Symposium and the 13th European
               Conference on Foundations of Software Engineering (ESEC/FSE)},
  pages     = {416--419},
  year      = {2011},
  doi       = {10.1145/2025113.2025179},
}

@article{panichella2018automated,
  author    = {Annibale Panichella and
               Fitsum Meshesha Kifetew and
               Paolo Tonella},
  title     = {Automated Test Case Generation as a Many-Objective Optimisation
               Problem with Dynamic Selection of the Targets},
  journal   = {IEEE Transactions on Software Engineering},
  volume    = {44},
  number    = {2},
  pages     = {122--158},
  year      = {2018},
  doi       = {10.1109/TSE.2017.2663435},
}

@inproceedings{cadar2008klee,
  author    = {Cristian Cadar and
               Daniel Dunbar and
               Dawson R. Engler},
  title     = {{KLEE}: Unassisted and Automatic Generation of High-Coverage Tests
               for Complex Systems Programs},
  booktitle = {Proceedings of the 8th USENIX Symposium on Operating Systems Design
               and Implementation (OSDI)},
  pages     = {209--224},
  year      = {2008},
}

@article{jia2011analysis,
  author    = {Yue Jia and
               Mark Harman},
  title     = {An Analysis and Survey of the Development of Mutation Testing},
  journal   = {IEEE Transactions on Software Engineering},
  volume    = {37},
  number    = {5},
  pages     = {649--678},
  year      = {2011},
  doi       = {10.1109/TSE.2010.62},
}

@inproceedings{coles2016pit,
  author    = {Henry Coles and
               Thomas Laurent and
               Christopher Henard and
               Mike Papadakis and
               Anthony Ventresque},
  title     = {{PIT}: A Practical Mutation Testing Tool for {Java}},
  booktitle = {Proceedings of the 25th International Symposium on Software Testing
               and Analysis (ISSTA)},
  pages     = {449--452},
  year      = {2016},
  doi       = {10.1145/2931037.2948707},
}

@inproceedings{niedermayr2016extreme,
  author    = {Rainer Niedermayr and
               Elmar Juergens and
               Stefan Wagner},
  title     = {Will My Tests Tell Me If I Break This Code?},
  booktitle = {Proceedings of the International Workshop on Continuous Software
               Evolution and Delivery (CSED)},
  pages     = {23--29},
  year      = {2016},
  doi       = {10.1145/2896941.2896944},
}

@article{veraPerez2018descartes,
  author    = {Oscar Luis Vera-P{\'{e}}rez and
               Benjamin Danglot and
               Martin Monperrus and
               Benoit Baudry},
  title     = {A Comprehensive Study of Pseudo-tested Methods},
  journal   = {Empirical Software Engineering},
  volume    = {24},
  number    = {3},
  pages     = {1195--1225},
  year      = {2019},
  doi       = {10.1007/s10664-018-9653-2},
}

@article{schuler2013checked,
  author    = {David Schuler and Andreas Zeller},
  title     = {Checked Coverage: An Indicator for Oracle Quality},
  journal   = {Software Testing, Verification and Reliability},
  volume    = {23},
  number    = {7},
  pages     = {531--551},
  year      = {2013},
  doi       = {10.1002/stvr.1497},
}

@inproceedings{just2014mutants,
  author    = {Ren{\'{e}} Just and
               Darioush Jalali and
               Laura Inozemtseva and
               Michael D. Ernst and
               Reid Holmes and
               Gordon Fraser},
  title     = {Are Mutants a Valid Substitute for Real Faults in Software Testing?},
  booktitle = {Proceedings of the 22nd ACM SIGSOFT International Symposium on
               Foundations of Software Engineering (FSE)},
  pages     = {654--665},
  year      = {2014},
  doi       = {10.1145/2635868.2635929},
}

@article{fraser2015achieving,
  author    = {Gordon Fraser and Andrea Arcuri},
  title     = {Achieving Scalable Mutation-based Generation of Whole Test Suites},
  journal   = {Empirical Software Engineering},
  volume    = {20},
  number    = {3},
  pages     = {783--812},
  year      = {2015},
  doi       = {10.1007/s10664-013-9299-z},
}

@book{ammann2016introduction,
  author    = {Paul Ammann and
               Jeff Offutt},
  title     = {Introduction to Software Testing},
  edition   = {2nd},
  publisher = {Cambridge University Press},
  year      = {2016},
  isbn      = {978-1107172012},
}

@book{osherove2014art,
  author    = {Roy Osherove},
  title     = {The Art of Unit Testing},
  edition   = {2nd},
  publisher = {Manning Publications},
  year      = {2014},
  isbn      = {978-1617290893},
}

@book{meszaros2007xunit,
  author    = {Gerard Meszaros},
  title     = {{xUnit} Test Patterns: Refactoring Test Code},
  publisher = {Addison-Wesley},
  year      = {2007},
  isbn      = {978-0131495050},
}

@book{myers2011art,
  author    = {Glenford J. Myers and
               Corey Sandler and
               Tom Badgett},
  title     = {The Art of Software Testing},
  edition   = {3rd},
  publisher = {John Wiley \& Sons},
  year      = {2011},
  isbn      = {978-1118031964},
}

@article{ostrand1988category,
  author    = {Thomas J. Ostrand and
               Marc J. Balcer},
  title     = {The Category-Partition Method for Specifying and Generating
               Functional Tests},
  journal   = {Communications of the ACM},
  volume    = {31},
  number    = {6},
  pages     = {676--686},
  year      = {1988},
  doi       = {10.1145/62959.62964},
}

@article{dakhel2024mutap,
  author    = {Arghavan Moradi Dakhel and
               Amin Nikanjam and
               Vahid Majdinasab and
               Foutse Khomh and
               Michel C. Desmarais},
  title     = {Effective Test Generation Using Pre-trained Large Language Models
               and Mutation Testing},
  journal   = {Information and Software Technology},
  volume    = {171},
  pages     = {107468},
  year      = {2024},
  doi       = {10.1016/j.infsof.2024.107468},
}

@article{yuan2024chattester,
  author    = {Zhiqiang Yuan and Mingwei Liu and Shiji Ding and Kaixin Wang and Yixuan Chen and Xin Peng and Yiling Lou},
  title     = {Evaluating and Improving {ChatGPT} for Unit Test Generation},
  journal   = {Proceedings of the ACM on Software Engineering},
  volume    = {1},
  number    = {FSE},
  pages     = {1703--1726},
  year      = {2024},
  doi       = {10.1145/3660783},
}

@article{landiskoch1977,
  author    = {J. Richard Landis and
               Gary G. Koch},
  title     = {The Measurement of Observer Agreement for Categorical Data},
  journal   = {Biometrics},
  volume    = {33},
  number    = {1},
  pages     = {159--174},
  year      = {1977},
  doi       = {10.2307/2529310},
}

@article{mccabe1976,
  author    = {McCabe, Thomas J.},
  title     = {A Complexity Measure},
  journal   = {IEEE Transactions on Software Engineering},
  volume    = {2},
  number    = {4},
  pages     = {308--320},
  year      = {1976},
  doi       = {10.1109/TSE.1976.233837},
}

@techreport{campbell2018cognitive,
  author      = {Campbell, G. Ann},
  title       = {Cognitive Complexity: A New Way of Measuring Understandability},
  institution = {SonarSource},
  year        = {2018},
  note        = {\url{https://www.sonarsource.com/docs/CognitiveComplexity.pdf}},
}

@article{barr2015oracle,
  author    = {Barr, Earl T. and Harman, Mark and McMinn, Phil and Shahbaz, Muzammil and Yoo, Shin},
  title     = {The Oracle Problem in Software Testing: A Survey},
  journal   = {IEEE Transactions on Software Engineering},
  volume    = {41},
  number    = {5},
  pages     = {507--525},
  year      = {2015},
  doi       = {10.1109/TSE.2014.2372785},
}

@inproceedings{andrews2005mutation,
  author    = {Andrews, James H. and Briand, Lionel C. and Labiche, Yvan},
  title     = {Is Mutation an Appropriate Tool for Testing Experiments?},
  booktitle = {Proceedings of the 27th International Conference on Software Engineering (ICSE)},
  pages     = {402--411},
  year      = {2005},
  doi       = {10.1145/1062455.1062530},
}

@article{chen2021codex,
  author    = {Chen, Mark and Tworek, Jerry and Jun, Heewoo and Yuan, Qiming and Pinto, Henrique Ponde de Oliveira and Kaplan, Jared and Edwards, Harrison and Burda, Yuri and Joseph, Nicholas and Brockman, Greg and others},
  title     = {Evaluating Large Language Models Trained on Code},
  journal   = {arXiv preprint arXiv:2107.03374},
  year      = {2021},
}

@article{harman2001sbse,
  author    = {Harman, Mark and Jones, Bryan F.},
  title     = {Search-based software engineering},
  journal   = {Information and Software Technology},
  volume    = {43},
  number    = {14},
  pages     = {833--839},
  year      = {2001},
  publisher = {Elsevier},
}

@inproceedings{pacheco2007randoop,
  author    = {Pacheco, Carlos and Lahiri, Shuvendu K. and Ernst, Michael D. and Ball, Thomas},
  title     = {Feedback-Directed Random Test Generation},
  booktitle = {Proceedings of the 29th International Conference on Software Engineering (ICSE)},
  pages     = {75--84},
  year      = {2007},
  organization = {IEEE},
}

@article{king1976symbolic,
  author    = {King, James C.},
  title     = {Symbolic Execution and Program Testing},
  journal   = {Communications of the ACM},
  volume    = {19},
  number    = {7},
  pages     = {385--394},
  year      = {1976},
  publisher = {ACM},
}

@article{demillo1978mutation,
  author    = {DeMillo, Richard A. and Lipton, Richard J. and Sayward, Frederick G.},
  title     = {Hints on Test Data Selection: Help for the Practicing Programmer},
  journal   = {Computer},
  volume    = {11},
  number    = {4},
  pages     = {34--41},
  year      = {1978},
  publisher = {IEEE},
}

@article{zhu1997software,
  author    = {Zhu, Hong and Hall, Patrick A. V. and May, John H. R.},
  title     = {Software Unit Test Coverage and Adequacy},
  journal   = {ACM Computing Surveys},
  volume    = {29},
  number    = {4},
  pages     = {366--427},
  year      = {1997},
  publisher = {ACM},
}

@inproceedings{just2014defects4j,
  author    = {Just, Ren{\'e} and Jalali, Darioush and Ernst, Michael D.},
  title     = {{Defects4J}: A Database of Existing Faults to Enable Controlled Testing Studies for {Java} Programs},
  booktitle = {Proceedings of the 2014 International Symposium on Software Testing and Analysis (ISSTA)},
  pages     = {437--440},
  year      = {2014},
  organization = {ACM},
}

@inproceedings{evalplus2023,
  author    = {Liu, Jiawei and Xia, Chunqiu Steven and Wang, Yuyao and Zhang, Lingming},
  title     = {Is Your Code Generated by {ChatGPT} Really Correct? {Rigorous} Evaluation of Large Language Models for Code Generation},
  booktitle = {Advances in Neural Information Processing Systems},
  volume    = {36},
  year      = {2023},
}

@inproceedings{alshahwan2024meta,
  author    = {Nadia Alshahwan and Jubin Chheda and Anastasia Finogenova and Beliz Gokkaya and Mark Harman and Inna Harper and Alexandru Marginean and Shubho Sengupta and Eddy Wang},
  title     = {Automated Unit Test Improvement using Large Language Models at {Meta}},
  booktitle = {Companion Proceedings of the 32nd ACM International Conference on the Foundations of Software Engineering (FSE)},
  pages     = {185--196},
  year      = {2024},
  doi       = {10.1145/3663529.3663839},
}

@inproceedings{lemieux2023codamosa,
  author    = {Lemieux, Caroline and Inala, Jeevana Priya and Lahiri, Shuvendu K. and Sen, Siddhartha},
  title     = {{CodaMosa}: Escaping Coverage Plateaus in Test Generation with Pre-trained Large Language Models},
  booktitle = {Proceedings of the 45th International Conference on Software Engineering (ICSE)},
  pages     = {919--931},
  year      = {2023},
  organization = {IEEE},
}

@article{goodenough1975toward,
  author    = {Goodenough, John B. and Gerhart, Susan L.},
  title     = {Toward a Theory of Test Data Selection},
  journal   = {IEEE Transactions on Software Engineering},
  volume    = {1},
  number    = {2},
  pages     = {156--173},
  year      = {1975},
  publisher = {IEEE},
}

@article{anand2013orchestrated,
  author    = {Anand, Saswat and Burke, Edmund K. and Chen, Tsong Yueh and Clark, John and Cohen, Myra B. and Grieskamp, Wolfgang and Harman, Mark and Harrold, Mary Jean and McMinn, Phil},
  title     = {An Orchestrated Survey of Methodologies for Automated Software Test Case Generation},
  journal   = {Journal of Systems and Software},
  volume    = {86},
  number    = {8},
  pages     = {1978--2001},
  year      = {2013},
  publisher = {Elsevier},
}

@inproceedings{beller2015developers,
  author    = {Beller, Moritz and Gousios, Georgios and Panichella, Annibale and Zaidman, Andy},
  title     = {When, How, and Why Developers (Do Not) Test in Their {IDE}s},
  booktitle = {Proceedings of the 10th Joint Meeting on Foundations of Software Engineering (ESEC/FSE)},
  pages     = {179--190},
  year      = {2015},
  organization = {ACM},
}

@inproceedings{shamshiri2015autogenerated,
  author    = {Shamshiri, Sina and Just, Ren{\'e} and Rojas, Jos{\'e} Miguel and Fraser, Gordon and McMinn, Phil and Arcuri, Andrea},
  title     = {Do Automatically Generated Unit Tests Find Real Faults? {An} Empirical Study of Effectiveness and Challenges ({T})},
  booktitle = {Proceedings of the 30th IEEE/ACM International Conference on Automated Software Engineering (ASE)},
  pages     = {201--211},
  year      = {2015},
  doi       = {10.1109/ASE.2015.86},
}

@inproceedings{shi2023detecting,
  author    = {Shi, Weijia and Ajith, Anirudh and Xia, Mengzhou and Huang, Yangsibo and Liu, Daogao and Blevins, Terra and Chen, Danqi and Zettlemoyer, Luke},
  title     = {Detecting Pretraining Data from Large Language Models},
  booktitle = {Proceedings of the 12th International Conference on Learning Representations (ICLR)},
  year      = {2024},
}

@incollection{offutt2001mutation,
  author    = {Offutt, A. Jefferson and Untch, Roland H.},
  title     = {Mutation 2000: Uniting the Orthogonal},
  booktitle = {Mutation Testing for the New Century},
  pages     = {34--44},
  year      = {2001},
  publisher = {Kluwer Academic Publishers},
}

@inproceedings{papadakis2018mutation,
  author    = {Mike Papadakis and Donghwan Shin and Shin Yoo and Doo-Hwan Bae},
  title     = {Are mutation scores correlated with real fault detection?: a large scale empirical study on the relationship between mutants and real faults},
  booktitle = {Proceedings of the 40th International Conference on Software Engineering (ICSE)},
  pages     = {537--548},
  year      = {2018},
  doi       = {10.1145/3180155.3180183},
}

@article{tip2025llmorpheus,
  author    = {Frank Tip and Jonathan Bell and Max Sch{\"a}fer},
  title     = {{LLMorpheus}: Mutation Testing Using Large Language Models},
  journal   = {IEEE Transactions on Software Engineering},
  volume    = {51},
  number    = {6},
  pages     = {1645--1665},
  year      = {2025},
  doi       = {10.1109/TSE.2025.3562025},
}

@inproceedings{hossain2023gaps,
  author    = {Soneya Binta Hossain and Matthew B. Dwyer and Sebastian G. Elbaum and Anh Nguyen-Tuong},
  title     = {Measuring and Mitigating Gaps in Structural Testing},
  booktitle = {Proceedings of the 45th International Conference on Software Engineering (ICSE)},
  pages     = {1712--1723},
  year      = {2023},
  doi       = {10.1109/ICSE48619.2023.00147},
}

@inproceedings{jahangirova2016oracle,
  author    = {Gunel Jahangirova and David Clark and Mark Harman and Paolo Tonella},
  title     = {Test oracle assessment and improvement},
  booktitle = {Proceedings of the 25th International Symposium on Software Testing and Analysis (ISSTA)},
  pages     = {247--258},
  year      = {2016},
  doi       = {10.1145/2931037.2931062},
}

@inproceedings{siddiq2024junit,
  author    = {Mohammed Latif Siddiq and Joanna Cecilia da Silva Santos and Ridwanul Hasan Tanvir and Noshin Ulfat and Fahmid Al Rifat and Vinicius Carvalho Lopes},
  title     = {Using Large Language Models to Generate {JUnit} Tests: An Empirical Study},
  booktitle = {Proceedings of the 28th International Conference on Evaluation and Assessment in Software Engineering (EASE)},
  pages     = {313--322},
  year      = {2024},
  doi       = {10.1145/3661167.3661216},
}

@article{tang2024chatgpt,
  author    = {Yutian Tang and Zhijie Liu and Zhichao Zhou and Xiapu Luo},
  title     = {{ChatGPT} vs {SBST}: A Comparative Assessment of Unit Test Suite Generation},
  journal   = {IEEE Transactions on Software Engineering},
  volume    = {50},
  number    = {6},
  pages     = {1340--1359},
  year      = {2024},
  doi       = {10.1109/TSE.2024.3382365},
}

@inproceedings{yang2024evaluation,
  author    = {Lin Yang and Chen Yang and Shutao Gao and Weijing Wang and Bo Wang and Qihao Zhu and Xiao Chu and Jianyi Zhou and Guangtai Liang and Qianxiang Wang and Junjie Chen},
  title     = {On the Evaluation of Large Language Models in Unit Test Generation},
  booktitle = {Proceedings of the 39th IEEE/ACM International Conference on Automated Software Engineering (ASE)},
  pages     = {1607--1619},
  year      = {2024},
  doi       = {10.1145/3691620.3695529},
}

@article{ryan2024codeaware,
  author    = {Gabriel Ryan and Siddhartha Jain and Mingyue Shang and Shiqi Wang and Xiaofei Ma and Murali Krishna Ramanathan and Baishakhi Ray},
  title     = {Code-Aware Prompting: A Study of Coverage-Guided Test Generation in Regression Setting using {LLM}},
  journal   = {Proceedings of the ACM on Software Engineering},
  volume    = {1},
  number    = {FSE},
  pages     = {951--971},
  year      = {2024},
  doi       = {10.1145/3643769},
}

@inproceedings{inozemtseva2014coverage,
  author    = {Laura Inozemtseva and Reid Holmes},
  title     = {Coverage is not strongly correlated with test suite effectiveness},
  booktitle = {Proceedings of the 36th International Conference on Software Engineering (ICSE)},
  pages     = {435--445},
  year      = {2014},
  doi       = {10.1145/2568225.2568271},
}

@article{gwet2008agreement,
  author  = {Kilem Li Gwet},
  title   = {Computing Inter-Rater Reliability and Its Variance
             in the Presence of High Agreement},
  journal = {British Journal of Mathematical and Statistical Psychology},
  volume  = {61},
  number  = {1},
  pages   = {29--48},
  year    = {2008},
  doi     = {10.1348/000711006X126600},
}
